\documentclass[sn-mathphys]{sn-jnl}

\jyear{2021}%

\theoremstyle{thmstyleone}%
\newtheorem{theorem}{Theorem}%  meant for continuous numbers
\newtheorem{proposition}[theorem]{Proposition}%

\theoremstyle{thmstyletwo}%
\newtheorem{example}[theorem]{Example}%
\newtheorem{remark}{Remark}%
\newcounter{assumption}
 \newtheorem{Assumption}[assumption]{Assumption}
 \newtheorem{Theorem}[theorem]{Theorem}

 \newcounter{lemma}
 \newtheorem{Lemma}[lemma]{Lemma}
 \newcounter{example}
 \newtheorem{Example}[example]{Example}
\theoremstyle{thmstylethree}%
\newtheorem{definition}{Definition}%
\usepackage{wasysym}
\usepackage{empheq}
\usepackage{url}
\usepackage{tabularx}
\usepackage{float,multirow,booktabs}\usepackage{array}
\usepackage[novbox]{pdfsync} \usepackage{relsize,exscale}
\usepackage[normalem]{ulem}
\usepackage{hyperref}
\usepackage{cleveref}
\usepackage{upgreek}
\usepackage{color}
\usepackage{xcolor}
\usepackage{subcaption}

\usepackage{graphicx}
\usepackage{epsfig}
\graphicspath{{figures/}}

\usepackage{geometry}

\usepackage{vmargin}
\setmarginsrb{.8in}{.6in}{.8in}{.4in}{0pt}{.3in}{0pt}{.3in}
\newcommand*{\QEDBL}{\hfill\ensuremath{\blacksquare}}
\def\Ity{It is easy to show that }
\newcommand*{\QEDB}{\hfill\ensuremath{\square}}%  empty square

\newcommand{\RN}[1]{\textup{\uppercase\expandafter{\romannumeral#1}}}

\def\Lgn{\L+\g+\nu_i}
\def\L{\Lambda}
\newcommand{\mL}{\mathcal{L}}\newcommand{\mV}{\mathcal{V}}

 \def\satd{satisfied}  \def\wmw{we may write}

\newcommand{\n}{\mathsf n}

\def\How{However, }\def\mbs{may be shown to  }\def\PF{Perron-Frobenius}
\def\no{\nonumber}\def\bep{\begin{pmatrix}} \def\eep{\end{pmatrix}}
 \def\DFE{disease free equilibrium}

\def\l{{\lambda}}
\newcommand{\map}[3]{#1: #2 \rightarrow #3}
\newcommand{\realpos}{\mathbb{R}_{> 0}}
\newcommand{\realnneg}{\mathbb{R}_{\geq 0}}
\newcommand{\naturalpos}{\mathbb{N}_{>0}}

\providecommand{\pp}[1]{\left[#1\right]} %[.]
\providecommand{\pr}[1]{\left(#1\right)} %(.)

\newtheorem{corollary}%[theorem]
{Corollary}
\def\beC{\begin{corollary}
  }\def\eeC{\end{corollary}}
  \def\beT{\begin{Theorem}}\def\eeT{\end{Theorem}}
\def\beA{\begin{Assumption}} \def\eeA{\end{Assumption}}
\def\beL{\begin{Lemma}} \def\eeL{\end{Lemma}}
\def\beXa{\begin{Example}} \def\eeXa{\end{Example}}
\def\wmc{we may check that }
{Conjecture}

\newcommand*{\QED}{\hfill\ensuremath{\blacksquare}}% Black square

\renewcommand{\theta}{\vartheta}

\renewcommand{\thefootnote}{\fnsymbol{footnote}}

\numberwithin{equation}{section}

\def\bc{\begin{cases}
  }      \def\mF{\mathcal F}
\def\ec{\end{cases}}  
  \def\qu{\quad} \def\for{\forall}

  \newcommand{\beq}{\begin{eqnarray}
    }
\def\eeq{\end{eqnarray}}
   \newcommand{\be}[1]{\begin{equation}\label{#1}}
\newcommand{\ee}{\end{equation}}
\def\bea{\begin{eqnarray*}}\def\ssec{\subsection}
  \def\Prf{{\bf Proof:} }  
\def\eea{\end{eqnarray*}}  \def\la{\label}\def\fe{for example }   \def\ith{it holds that } 
   \def\Mp{More precisely, } \def\sats{satisfies}  \def\saty{satisfy}       \def\Itm{It may be checked that }\def\Oth{On the other hand, }
\def\sd{s_{dfe}} \def\rd{r_{dfe}}
\def\I{\infty} \def\Eq{\Leftrightarrow}

\def\BEN{\begin{enumerate}}  \def\BI{\begin{itemize}}
\def\EEN{\end{enumerate}}   \def\EI{\end{itemize}} \def\im{\item} \def\Lra{\Longrightarrow}  \def\eqr{\eqref}  
\def\no{\nonumber} 
 \def\nR{\mathcal R}
\def\mR{\mathcal R_0} \def\oth{otherwise}
\def\g{\gamma}     \def\b{\beta}

\def\Fr{Furthermore, }\def\wms{we must show that }
    
   \def\Mp{More precisely, } 
   \def\wrt{with respect to }
  \def\Equ{Equivalently, }\def\resp{respectively}  \def\fno{from now on} 
 \def\eqr{\eqref}  
\def\wk{well-known}\def\itm{it may be shown that }

\def\l{\; \mathsf l}
\def\fr{\frac} \def\im{\item}
\newcommand{\s}{\;\mathsf s}
\renewcommand{\i}{\;\mathsf i}
\renewcommand{\r}{\; \mathsf r}\def\m0{{\mathcal R}_0}

\def\eeD{\end{definition}} \def\beD{\begin{definition}}
\def\beR{\begin{remark}} \def\eeR{\end{remark}}
\def\mD{\mathcal D }
         
 \def\vi{\vec \i}  \def\vz{\vec 0}  \def\sd{s_{dfe}}

    \def\beP{\begin{proposition}} \def\eeP{\end{proposition}}
      
    \def\it{immunity threshold}
    
    \long\def\symbolfootnote[#1]#2{
\begingroup
\def\thefootnote{\fnsymbol{footnote}}\footnote[#1]{#2}
\endgroup}
\def\fn{\symbolfootnote}

\def\l{\lambda}%\def\e{i_{ref}}
\def\com{compartment}
\def\D{\Delta}

\def\Itf{It follows that } 
   \def\frt{furthermore }
\def\brn{basic reproduction number}\def\ngm{next generation matrix}
\newcommand{\red}{\textcolor[rgb]{1.00,0.00,0.00}}
\newcommand{\blue}{\textcolor[rgb]{0.00,0.00,1.00}}
\newcommand{\figu}[3]{
\begin{figure}[H]
\centering
\includegraphics[width=\textwidth]{#1}
\caption{#2\label{f:#1}}
\end{figure}
}
\renewcommand{\(}{\left(}
\renewcommand{\)}{\right)}

\begin{document}

\title[ ]{Stability analysis of an eight parameter SIR-type  model including loss of immunity, and disease and vaccination fatalities}


\author*[1]{\fnm{Florin} \sur{Avram}}\email{florin.avram@univ-pau.fr}

\author[2]{\fnm{Rim} \sur{Adenane}}\email{rim.adenane@uit.ac.ma}
\equalcont{These authors contributed equally to this work.}

\author[3]{\fnm{Gianluca} \sur{Bianchin}}\email{gianluca.bianchin@colorado.edu}
\equalcont{These authors contributed equally to this work.}

\author[4]{\fnm{Andrei  } \sur{Halanay}}\email{andrei.halanay@upb.ro}
\equalcont{These authors contributed equally to this work.}

\affil*[1]{\orgdiv{Laboratoire de Math\'ematiques Appliqu\'es}, \orgname{Universit\'e de Pau}, \orgaddress{ \city{ Pau}, \postcode{64000},  \country{France}}}

\affil[2]{\orgdiv{Department of Mathematics}, \orgname{Ibn-Tofail University}, \orgaddress{ \city{Kenitra}, \postcode{14000},  \country{Morocco}}}

\affil[3]{\orgdiv{Department of Electrical, Computer, and Energy Engineering}, \orgname{University of Colorado Boulder}, \orgaddress{ \city{Boulder}, \postcode{CO 80309},  \country{United States}}}

\affil[4]{\orgdiv{Department of mathematics and informatics}, \orgname{ Polytechnic University of Bucharest}, \orgaddress{ \city{Bucharest},  \country{Romania}}}


\abstract{We revisit here  a landmark five parameter SIR-type model of \cite[Sec. 4]{Der}, which is maybe the simplest  example where a complete picture of all cases, including non-trivial bistability behavior,  may be obtained using  simple tools.
We also generalize it by  adding  essential vaccination and  vaccination induced death parameters, with the aim of revealing the role of  vaccination and its possible failure. The main result is Theorem \ref{cases}, which describes the stability behavior of our model in all possible cases.
}


\maketitle

\tableofcontents
\section{Introduction}\label{sec1}
{\bf Motivation}. This paper has the dual purpose of providing  a short guide to students of deterministic mathematical epidemiology,   among which we count ourselves, and at the same time to illustrate the technical work one is faced with in an elementary, but not  simple ``exercise". Of course, one can easily find at least five must-read excellent textbooks and thesis surveying this field (with different emphasis: epidemiological, stability, or control), see for example~\cite{anderson1992infectious,smith2011dynamical,Mart,Thieme,Chavez,mondaini2020trends,bacaer2021mathematiques,
della2021problemi}, and also at least a hundred major papers which are a  must read.  We hope however that our little guide may help future students decide as to the order in which these materials must be assimilated.

{\bf A bit of history}. Deterministic
mathematical modelling  of diseases started with works of Ross on malaria, and imposed itself
after   the  work of \cite{KeMcK} on the Bombay plague of 1905--06. This was subsequently followed by works    on measles,
smallpox,chickenpox, mumps, typhoid fever and diphtheria, see \fe~\cite{earn2008light}, and~recently  the COVID-19 pandemic (see \fe\  \cite{Schaback,bacaer2020modele,Ketch,Charp,Djidjou,Sofonea,alvarez2020simple,
horstmeyer2020balancing,
di2020optimal,Franco,baker2020reactive,caulkins2020long,caulkins2021optimal},
to~cite just a few representatives of a huge literature). 
Note that at its beginning, mathematical epidemiology was a collection of similar examples dealing with current epidemics, and no one bothered
to define precisely what is an epidemic, mathematically. One answer to this question, reviewed in Appendix B, may be found in the recent paper \cite{KamSal}.

{\bf The deterministic mathematical epidemiology literature} may be divided  into three streams.
\BEN \im ``Constant total population'' models 
 are the easiest  to study.
 \How\ since death is an essential factor of epidemics, the assumption of constant total population  (clearly a short term or very large population approximation)  deserves some  comment. {One possible rigorous justification for deterministic constant population  epidemiological models comes from slow-fast analysis  \cite{KK,Kuehn,Ginoux}.   This is best understood for models with demography (birth,  death), which happen typically on a  slower scale than the infectious  phenomena. Here there is a natural partition of the compartments
into a vector $\vi(t) \in \mathbb{R}_+^-$ of disease/infectious \com s (asymptomatic, infectious hospitalized,etc).  These interact (quickly) with the other input classes like the susceptibles and output classes like the recovered and dead.} 

 For constant total population'' models  the total population $N$ clearly plays no role, and one may use the assumption that the rate of infection is independent of  $N$,  of the form $\b S(t) I(t)$ (which was called pseudo, or simple mass-action incidence  \cite{de1995does}). Note that this simplification of the rate of infection (adopted in the majority of the literature) is  inappropriate for varying population models.

\im Models with constant birth rate (in the analog stochastic model, this would correspond to emigration). These models include the previous class, and preserve some of its nice features, like the uniqueness of the endemic fixed point. They  typically   \saty\ the ``$\m0$ alternative'', established  via the  \ngm\ approach, and also the fact that the endemic point exists only when $\m0>1$ -- see
\cite{de2019some} for a recent  nice review of several stability results for  this class  of models.

\im  Finally, we arrive to the class our paper is concerned with:  models with   linear birth-rate $\L N$ ( {or constant birth-rate per capita in the analog stochastic model}), varying total population $N$, and ``proportionate
mixing'' rate of infection $$\b S(t) \fr{I(t)}{N(t)}.$$ {As far as we know, this stream of literature  was initiated in \cite{busenberg1990analysis,busenberg1993method}, and  allows the possibility of bi-stability when $\m0 <1$ (absent from the previous models), even in a  SIR example with five parameters} \cite{Der}.\fn[5]{This reveals that for an initial high
number of infectives, the trajectory may  lie in the
basin of attraction  of a stable
endemic  point instead of being eradicated. The discrepancy with what is expected from the corresponding stochastic model suggests that in this range the deterministic model is inappropriate}.

This last stream of literature is quite important, since constant birth-rate per capita is a natural assumption.

\EEN

Despite further remarkable works  on particular examples -- see for example \cite{Green97,LiGraef,Raz,li2002qualitative,SunHsieh,Arinovar,lu2018global%,shi2012dynamical
} ({which preferred all direct stability analyses to  the \ngm\ approach}), the  literature on models with varying total population, unlike the  two preceding streams, has not yet reached  general  results.

To understand  this failure, it seemed  to us  a good idea to revisit an  important   SIR model with disease-induced deaths  and  loss of immunity~\cite{Der,Raz}. These important  works   illustrate already some of  the complexities which may arise for varying population models, in particular the possibility of bi-stability when $\m0 <1$.
This surprising fact %that we cannot expect in general global stability for  this class
lead us to introduce the  concept of strong global  stability of the DFE (\DFE), and to  find conditions for this to hold in our example, (see Proposition \ref{p:sgs}). Since the method used  is simply  linear programming, we hope to extend  this in future work.

{\bf Our model}. We further added to the extension of  \cite{KeMcK} introduced in ~\cite{Der,Raz} {vaccinations}.
Grouping together the recovered and vaccinated,  our ``SIR/V+S'' model  is described by:
\begin{align}
\la{SIRNe}
S'(t)&=\L N(t) -\fr{\b}{N(t)} S (t) I (t) -(\g_s+ \mu) S(t) + \g_r R(t),
\nonumber\\
I'(t)&= I(t)\pr{\fr{\b}{N(t)}  S(t)+\fr{\b_r}{N(t)} R (t)-\g   -\mu-\nu_i},\nonumber\\
R'(t)&= \g_s S(t)+\g    I(t) -(\g_r+\mu+\nu_r) R(t)-\fr{\b_r}{N(t)} R (t)
I (t), \nonumber\\
D'(t)&= \mu(   S(t)+ I(t)+R(t)),\nonumber\\
D_e'(t)&=\nu_i  I(t)+ \nu_r    R(t),\nonumber \\
N'(t)&=(S(t)+ I(t)+R(t))'=(\L  -  \mu) N(t) -\nu_i  I(t) -\nu_r  R(t),
\end{align}

It involves six  states:
$\map{S}{\realnneg}{\naturalpos}$ describing the number of
susceptible individuals in the population,
$\map{I}{\realnneg}{\naturalpos}$ describing the number of
infections,
$\map{R}{\realnneg}{\naturalpos}$ describing the number of recovered
or vaccinated,
$\map{D}{\realnneg}{\naturalpos}$ describing the number of natural
deaths in  the population,
$\map{D_e}{\realnneg}{\naturalpos}$ describing the number of deaths
originated by the disease, and
$\map{N}{\realnneg}{\naturalpos}$ describing the total number of
(alive) individuals in the population.

The parameters $\L \in \realnneg$  and $ \mu \in \realnneg$ denote the average
birth and death rates in the population (in the absence of
the disease), respectively,
$\g_s \in \realnneg$  is  the  vaccination rate,
$\g_r \in \realnneg$  denotes the rate at which immune individuals
lose immunity (this is the reciprocal of the expected duration of
immunity),  $\g  \in \realpos$ is the  rate at which infected
individuals recover from the disease,
 $\nu_i \in \realnneg$  is the
extra death rate due to the disease,  and $\nu_r \in \realnneg$  is
an average death rate in  the recovered-vaccinated \com~(due to e.g.
deaths caused by the vaccine). Note that in what follows we use the notation
$\g_c$ to denote the total rate at which individuals leave a certain
compartment $C$ towards other  non-deceased compartments, and we use
$\nu_c$ to denote the rate at which individuals leave compartment $C$
towards a deceased compartment.

In \eqref{SIRNe}, susceptible individuals become infected
at rate $\frac{\beta}{N(t)}I(t)$ (thus moving to the $I$
compartment), they are vaccinated at rate
$\gamma_s$ (thus moving to the $R$ compartment), and deaths occur
at rate $\mu$ (thus moving to the $D$ compartment); infected
individuals recover at rate $\gamma$ (thus moving to the $R$
compartment), die of non-disease related causes at rate $\mu$ (thus
moving to the $D$ compartment), and die of disease-related causes at
rate $\nu_i$ (thus moving to the $D_e$ compartment); recovered
individuals lose immunity at rate $\gamma_r$ (thus moving to the $S$
compartment), die of non-disease related causes at rate $\mu$ (thus
moving to the $D$ compartment), die of disease-related causes at rate
$\nu_r$, and become re-infected at rate $\fr{\b_r}{N(t)}I (t)$ (thus
moving to the $I$ compartment) (see Remark~\ref{r:super_infections}
for a discussion  on re-infections).

Note that  ${D,D_e}$   are completely determined once the other classes
are found. These ``output classes''  will not be mentioned further (since they are only relevant in control problems, which are outside our scope). The dynamics of $I$,  the disease class, allows computing the \brn\ via the \ngm\ method. Finally, the  {input classes} $S,R$  determine by themselves the \DFE.

\beR \la{r:super_infections} {Models of the form \eqref{SIRNe} that account for
non-constant population sizes can be especially useful in two
scenarios: (i) to study diseases that remain infectious for long
periods of time with small disease mortality rate, where the natural
death/birth rate of the population plays a central role (such as
HIV/AIDS, malaria and tuberculosis), as well as (ii) to study
diseases with short infectious periods but with a substantial disease
mortality rate, where the death rate due to the disease plays a
central role (such as measles, influenza, SARS/COVID).}

The  two-way transfers between the
recovered and infected compartments (recall that
$R\overset{\beta_r}{\underset{\gamma}{\rightleftarrows}}I$).
 can be used to account for multiple variants of the
disease, whereby immunity to one variant does not guarantee immunity
to all other variants. For instance, diseases such as HIV can develop
resistance to medications, and such resistance can be transmitted
to a partner. In this cases, even when the second party has
recovered, it may become re-infected with a different variant.
Notice that, an even more general case is considered in \cite{Raz},
where vaccinated individuals may transition to the infected
compartment.
\eeR

\beR \la{r:par_values}
In practical situations, certain  parameter relations,  for instance
$\b_r >\b, \nu_r >\nu_i,$ might seem ``unreasonable from a medical point of view''.
In what follows, we choose not to assume any relationship among the
parameters in \eqref{SIRNe} in order to highlight the fact surprising
mathematical behaviors like bistability-- see Figure~\ref{fig:2endpts} -- may arise when ``things go wrong".
\eeR

We make now several remarks, which serve as an appetizer for the rest of the paper.

 \beR \la{r:brz}

 The critical value $\b=\b_r$ defines a model where both the sickness and the vaccination do not affect the infectivity (neither confers  any immunity). {The recovered class might be better viewed then as a susceptible  class $S_o$ of older individuals, with extra mortality rate $\nu_r >0$ and no births}:
\begin{align*}
%\la{SIRNec}
S'(t)&=\L N(t) -\fr{\b}{N(t)} S (t) I (t) -(\g_s+ \mu) S(t) + \g_r S_o(t),
\nonumber\\
S_o'(t)&= -\fr{\b}{N(t)} S_o (t)
I (t) -(\g_r+\mu+\nu_r) S_o(t)+\g_s S(t)+\g    I(t),\\
I'(t)&= I(t)\pr{\fr{\b}{N(t)} \pr{ S(t)+S_o (t)}-\g   -\mu-\nu_i}.\nonumber
\end{align*}

A moment of reflection reveals that this particular case has two surprising  features: a) after the infection is over, transfers only occur towards the old class $S_o$,  and  b) transfers between the two age groups occur; it is hard to make sense of this, without further imposing $\g_s=\g_r=0$.

\eeR

\beR When $\b_r=0$, our model is an  example of matrix-SIR Arino model with linear birth rate, a class of models for which only few general results are available \cite{AABBGH}. \eeR

In what follows, we will investigate the stability properties of an
equivalent system to \eqref{SIRNe} obtained by investigating the normalized
quantities
$%\label{eq:fractionss}
\s=\fr{S}N,
\i=\fr{I}N,
\r=\fr{R}N$ and given by (see Proposition \ref{p:_model}, Section \ref{s:dim} for a detailed derivation).
\bea
\bc
&\s'(t) =\L + \g_r \r(t) -  \pr{\b \i(t)+\g_s} \s(t)  +\s(t)(-\L+ \nu_i \i(t)+\nu_r \r(t)),\\
&\i'(t) =\i(t)\pp{\b \s(t) +\nu_i  \i(t)+(\nu_r +\b_r) \r(t) - \pr{\g  +{\nu_i  +\L}}},\\
&\r'(t) = \g   \i(t) +\g_s \s(t) -{(\g_r+\L+\nu_r)} \r(t)+\r(t)(-\b_r \i(t)+\nu_i \i(t)+\nu_r \r(t)),\\
&\s(t)+\i(t)+\r(t)=1.
\ec
\eea

\beR Note this reduces to the classic SIR model \cite{KeMcK} when $\L=\nu_i=\nu_r=\g_r=\g_s=\b_r=0$. \eeR

\beR
Factoring the second equation reveals the threshold after which the infection starts decreasing
\be{il} \b \s(t) +\nu_i  \i(t)+(\nu_r +\b_r) \r(t) <  \Lgn \ee
{\bf  herd immunity $(\s,\i)$ line}, or max-line (line since $\r=1-\s-\i$). Notice that, in contrast with the case of models with constant
population size and no loss of immunity where the herd immunity
condition depends only on the susceptible state, the above condition
depends on three states $(\s,\i, \r)$. Also,  when $\nu_i=\nu_r=\b_r=0$, this
reduces to the well-studied herd immunity threshold.

1)  The inequality obtained at the DFE, when $\i=0$, \be{R1}
 \b \sd +(\nu_r +\b_r) \rd \leq  \Lgn \Eq \fr{\b \sd +(\nu_r +\b_r) \rd}{\Lgn}:=\m0 <  1\ee
  turns out
 to ensure  the local stability of the disease free equilibrium (\DFE)\   -- see section \ref{s:ngm}.

2) $\m0$ introduced above coincides with the famous \brn\ defined via the \ngm\ approach.

3) The disease free equilibrium (obtained by plugging $\i=0,\r=1-\s$ in the fixed point equation) is such that
\be{sdfe}
 s_{dfe}:=  \fr{\L +\g_r}{\L+\g_r+\g_s} \in [0,1], \quad s_{dfe}=1 \Eq \g_s=0
\ee

 \eeR

 \beR \la{r:gs}   A) The equality  $  \m0 = 1$ is linear in $\g_s$ and yields the  so called
 ``critical vaccination''%Gene-case.nb
\be{crv}   {\g_s^*}=
\pr{\Lambda +\gamma _r} \frac{\beta-\pr{\Lgn} }{\Lgn-\beta _r}, \ee
provided that the denominator doesn't blow-up.

This formula is positive if either $\b_r \le \Lgn \le \b$, or $\b \le \Lgn \le \b_r$.
When $\b_r=0$, we recover a classical critical vaccination formula

\be{nR}{\g_s^*}=\pr{\Lambda +\gamma _r}\(\nR-1\), \nR=\frac{\beta }{\Lgn}.\ee

B) The equality  $  \m0 = 1$ is also linear in $\b$ and yields a
 ``critical contact rate "
 \be{bc} \b^*= \frac{\left(\gamma +\nu _i+\Lambda \right) \left(\Lambda +\gamma _r+\gamma _s\right)-\beta _r \gamma _s}{\Lambda +\gamma _r}.\ee
 \Itm at this critical value the value $i_{ee}$ of the infectious at the lower endemic point  crosses the $i=0$ axis, and may reduce the  number of endemic points from 3 to 2  -- see Figure \ref{f:bif} for details.
\eeR

{\bf  Contents}.
Section \ref{s:dim} reviews the  dimension reduction available for the
proportions of models with linear birth rate, and emphasizes the fact
that the well studied deterministic model is an approximation of the
model with linear birth rate studied here. A second, finer
``intermediate approximation'' is introduced as well.

Section \ref{s:ngm} computes $\m0$ via the \ngm\ approach (NGM), establishing thereby the \wk\ weak $\m0$ alternative \cite{Van08}; it also introduces the  concept of ``strong global  stability'' of the DFE in Proposition \ref{p:sgs}, which may be useful for more advanced~models.

The endemic equilibria are discussed in Section \ref{s:end}.

Section \ref{s:nr0} identifies more precisely,  in the particular case $\nu_r=0$,  the case when global  stability does not hold.  The results involve  heavily the  vaccination parameter $\g_s$ and its critical value. The increased complexity of the model forces a geometric approach, already hinted at in \cite[Sec. 4]{Der}. This ends up  in the consideration of 10 cases, one of which, Theorem \ref{cases}2.(c), remains only partly resolved.

Section \ref{s:SIRI} discusses the particular case $\g_s=0$, generalizing and providing missing details of the results in \cite[Sec. 4]{Der}.

 Section \ref{s:FOA}  gives  the simpler results for  the  first approximation FA (actually  for a slightly more general ``classic/pedagogical model'').

 Finally, Section \ref{s:lem} provides the proof of a technical result, and Section \ref{s:pr} reviews  the pillar of deterministic epidemic models:
the definition of the \brn\ via the \ngm\ method.

\section{Dimension reduction  for the SIR/V+S model with linear birth rate}\label{s:dim}

It is  convenient to reformulate \eqr{SIRNe} in terms of the
normalized fractions
\begin{align}
\label{eq:fractionss}
\s&=\fr{S}N, &
\i&=\fr{I}N,&
\r&=\fr{R}N.
\end{align}
This process, sometimes called ``non-dimensionalizing'' (see
\fe\ \cite{rashkov2021model}),  allows us to provide the
following equivalent representation of \eqr{SIRNe}.

\beP \label{p:_model}
Let $\s, \i, \r$ be as defined in \eqref{eq:fractionss}. Then, the
dynamics \eqref{SIRNe} can be equivalently rewritten as:
\begin{align}
\label{SIRsc}
\s'(t) &=
\L + \g_r \r(t) -  \pr{\b \i(t)+\g_s} \s(t)  +\s(t)(-\L+ \nu_i \i(t)+\nu_r \r(t)), \nonumber\\
\i'(t) &=%\frac{I'(t)}{N}=
 \i(t)\pp{\b \s(t) +\nu_i  \i(t)+(\nu_r +\b_r) \r(t) - \pr{\g  +{\nu_i  +\L}}}, \nonumber\\
\r'(t) &= \g   \i(t) +\g_s \s(t) -{(\g_r+\L+\nu_r)} \r(t)+\r(t)(-\b_r \i(t)+\nu_i \i(t)+\nu_r \r(t)),\nonumber\\
\s(t)&+\i(t)+\r(t)=1.
\end{align}
\eeP

\begin{proof}
By using \be{N}N'(t)/N(t)=(\L  -\mu)-\nu_i  \i(t)-\nu_r \r(t),\ee we
obtain for the susceptibles:
\begin{align*}
\s'(t)&= \frac{S'}{N}- \frac{N'}{N^2}S \nonumber\\
& = \frac{\L(N)}{N}- \frac{\b S(t) I(t)}{N^2} + \frac{\g_r R(t)-(\mu +\g_s) S(t)}{N}- S(t) \frac{(\L  -\mu)-\nu_i  \i(t)-\nu_r \r(t)  }{N} \nonumber\\
&= \L(1-\s(t)) -\b \s(t)\i(t) + \g_r \r(t)+ \s(t)\left(\mu+\nu_i  \i(t)+\nu_r \r(t)-\mu  -\g_s\right)\nonumber\\
&=  \L -\b  \s(t)\i(t) + \g_r \r(t)- {(\g_s +  \L)}\s(t)+\s(t)(\nu_i \i(t)+\nu_r \r(t)).
\end{align*}

Similarly,
\begin{align*}
\i'(t)&= \frac{I'}{N}- \frac{N'}{N^2} I =  \b \s(t) \i(t) - (\g   +\mu+\nu_i) \i(t) -I(t) \frac{(\L  -\mu)-\nu_i  \i(t)-\nu_r \r(t) }{N}\\
&=  - \i(t) \L +\i(t)(\b \s(t) +\b_r \r(t) ) - (\g   +\mu+\nu_i) \i(t)  + \i(t)\left(\mu +\nu_i  \i(t)+\nu_r \r(t) \right)\\
&=  \i(t)\pp{\b \s(t) +\nu_i  \i(t) +(\b_r  +\nu_r) \r(t) -  \pr{\Lgn }},
\end{align*}
and
\begin{align*}
\r'(t)&= \frac{R'}{N}- \frac{N'}{N^2}R \\
&= \g   \i(t) +\g_s \s(t) - ( \g_r +\mu+\nu_r) \r(t)- R(t) \frac{(\L  -\mu)-\nu_i  \i(t)-\nu_r \r(t)  }{N} \nonumber
\\&= \g   \i(t) +\g_s \s(t) -{(\g_r+\L+\nu_r)} \r(t)+\r(t)(-\b_r \i(t)+\nu_i \i(t)+\nu_r \r(t)).
\end{align*}
Finally, $\s(t)+\i(t)+\r(t)=1$ follows from
$N'(t)=(S(t)+ I(t)+R(t))'$ by substituting  \eqref{eq:fractionss},
which proves the equivalence between \eqref{SIRNe} and \eqref{SIRsc}.
\QEDBL
\end{proof}

\beR Note that  the natural death rate $\mu$  does not intervene in  \eqref{SIRsc}, which is to be expected. Indeed,
 since this rate is the same
for all the compartments, it has no effect on the fractions.\eeR

\beR
\BEN
\im Note that the conservation equation $$\n:=\s+\i+\r=1,$$ in general,
does not follow from the first three equations in \eqref{SIRsc}.
Indeed, by summing the right hand-sides  we have:
$$\n'(t)= \pp{1-\s(t)-\r(t)-\i(t)}(\Lambda -\nu_i \i(t)-\nu_r \r(t)),$$
which shows that $\n'(t) \neq 0$ in general cases. However, the above
differential equation guarantees that  if
$\s(t_0)+\i(t_0)+\r(t_0)=1$ for some $t_0 \in \realnneg$, then
$\s(t)+\i(t)+\r(t)=1$ for all $t \geq t_0$. Accordingly, the
manifold
$$ \mD:=\{\s+ \i+\r=1,\s\geq 0, \i \geq 0, \r \geq 0\}$$ is
forward-invariant, since the flow along its boundaries is directed
towards the interior -- see \cite{Raz}\fn[4]{This reduction  of the
state space is important, since \oth\  we get an additional {fixed
point} obtained from the first factor above which turns out not to
satisfy the conservation equation,  and  makes no sense from  the
epidemiologic point of view.}. Note that the conservation equation
may replace either the last or the first equation in the dynamics,
and  allows reducing the computation of fixed points to dimension
$2$.

\im
The Jacobian matrix of \eqr{SIRsc} is given by
\be{Ja}
J(s,i)=  \left(
\begin{array}{cc}
 i \left(\nu _i-\beta \right)-\Lambda -\gamma _r-\gamma _s & s \left(\nu _i-\beta \right)-\gamma _r \\
 i \left(\beta -\beta _r\right) & -\gamma +2 i \nu _i-\nu _i-2 i \beta _r-\Lambda +\beta _r+s \left(\beta -\beta _r\right) \\
\end{array}
\right).
\ee
Following \cite{LiGraef}, let us investigate when this system is order preserving in the interior of the convex feasible region $\mD$.  The Jacobian \eqr{Ja} is a Metzler matrix $\for (s,i)\in \mD$ iff   $\nu_i-\b -\g_r > 0, \g_r \leq 0, \b>\b_r\Eq \nu_i>\b+\g_r , \g_r\leq 0, \b>\b_r$.
Thus, if $\bc  \nu_i>\b+\g_r \\ \g_r\leq 0\\ \b>\b_r\ec,$ the system \eqr{SIRsc} is order preserving  \wrt\ the partial ordering defined by the orthant $\left\{   (s,i)\in \mathbb{R}^2,\; s\geq 0,\; i\geq 0\right\}$.
\begin{figure}[H] %DerJustification.nb
    \centering
        \includegraphics[scale=0.7]{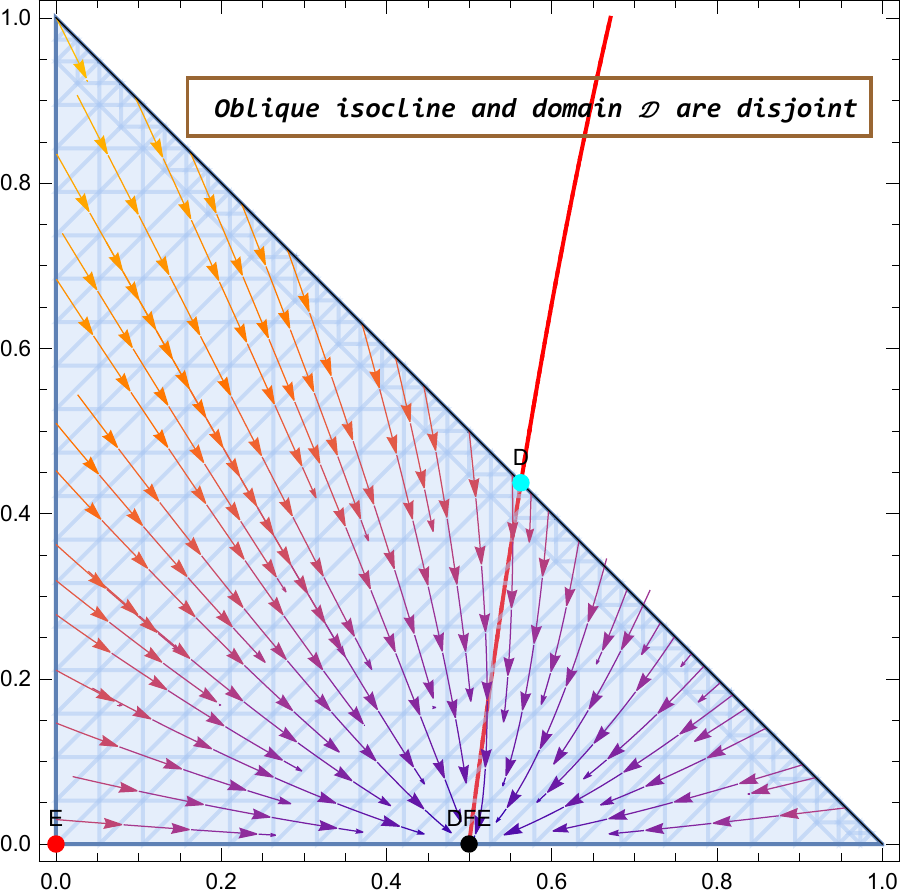}
        \caption{Stream Plot of $(\s,\i)$ when $\nu_i>\b+\g_r , \g_r= 0, \b>\b_r$, with $\beta = \frac{1}{2},\b_r= \frac{1}{4},\Lambda = 1,\nu_i= 1.01,\gamma= 1,\g_r= 0,\g_s= 1$. The DFE is a stable sink.}
        \label{fig:MonoS1}
    \end{figure}
Analogously, if $\bc \nu_i<\b+\g_r \\ \g_r>0\\ \b<\b_r\ec,$ the system \eqr{SIRsc} is order preserving  \wrt\ the partial ordering defined by the orthant $\left\{   (s,i)\in \mathbb{R}^2,\; s\leq 0,\; i\geq 0\right\}$.
\begin{figure}[H] %DerJustification.nb
    \centering
    \begin{subfigure}[a]{0.30\textwidth}
        \includegraphics[width=\textwidth]{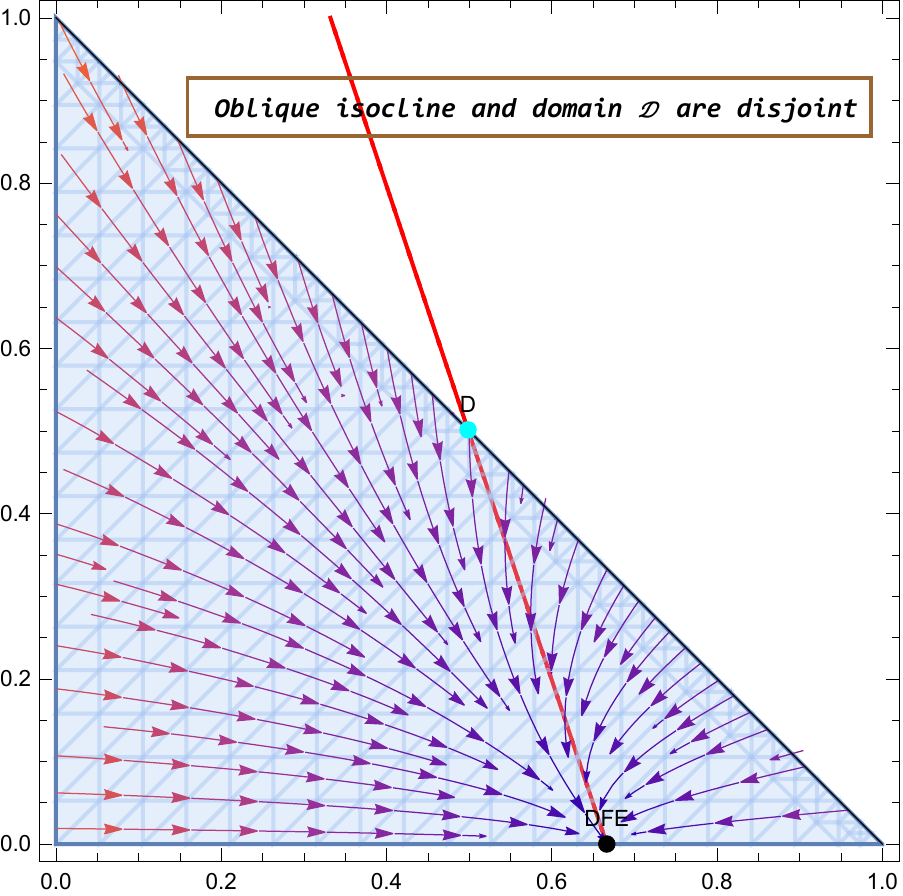}
        \caption{$\m0<1$, with $\beta = 1.01,\beta _r= 2,\Lambda = 1,\nu _i= 1,\gamma = 1,\gamma _r= 1,\gamma _s= 1$, the DFE is a stable sink.}
        \label{fig:MonoS2}
    \end{subfigure}
     ~
    \begin{subfigure}[a]{0.30\textwidth}
        \includegraphics[width=\textwidth]{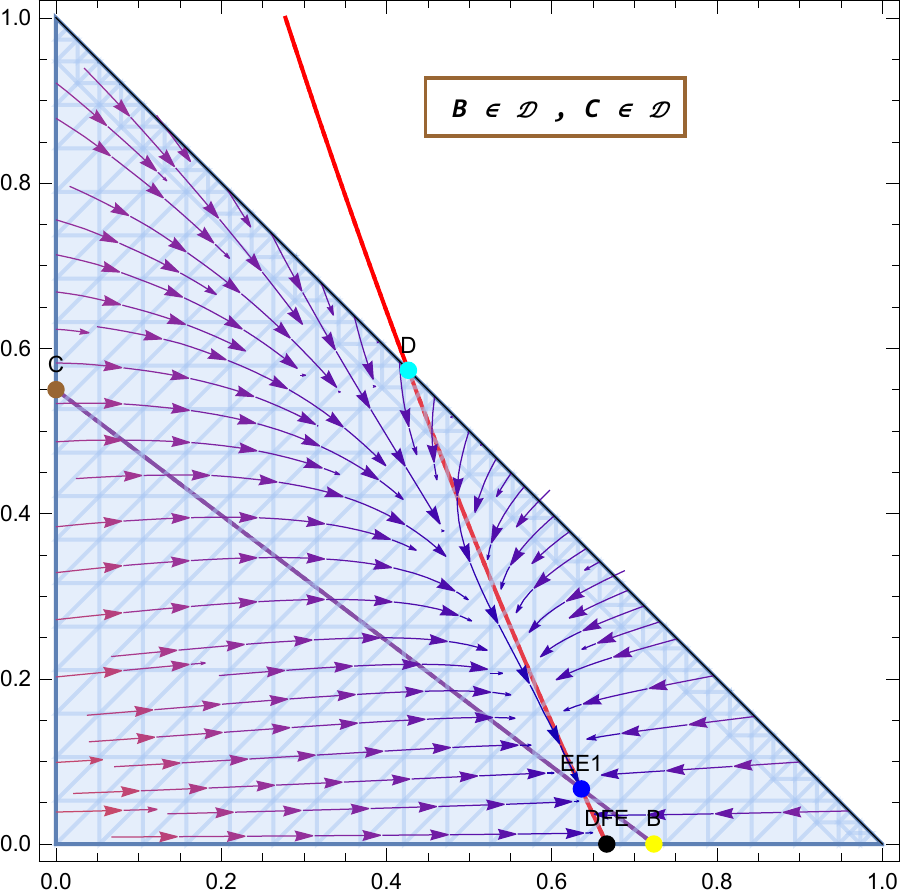}
        \caption{$B, C\in \mD, \m0>1$, with $\beta = 1.1,\beta _r= 3,\Lambda = 1,\nu _i= \frac{1}{2},\gamma = \frac{1}{8},\gamma _r= 1,\gamma _s= 1$, The DFE is a saddle point and EE is a stable sink.  }
        \label{fig:MonoB}
    \end{subfigure}
     \caption{Stream Plot of $(\s,\i)$ when $\nu_i<\b+\g_r , \g_r>0, \b<\b_r$.}
     \label{fig:Mono}
    \end{figure}

\EEN
\eeR

The study of the dynamics in Proposition~\ref{p:_model} is  quite challenging, and it may be useful sometimes to consider also the two approximations introduced in  the
following definition.

\beD \la{d:fis} Let $\Phi_s, \Phi_i, \Phi_r \in \{0,1\}$ and let
\begin{align}
\label{SIRscD}
\s'(t) &=
\L -\b  \s(t)\i(t) + \g_r \r(t)- {(\g_s +  \L)}\s(t)+\Phi_s\s(t)(\nu_i \i(t)+\nu_r \r(t)), \nonumber\\
\i'(t) &=
 \i(t)\pp{\b \s(t) +\b_r \r(t) - \pr{\g  +{\nu_i  +\L}}}
 + \Phi_i \i(t)(\nu_i  \i(t) + \nu_r \r(t)), \nonumber\\
\r'(t) &= \g   \i(t) +\g_s \s(t) -{(\g_r+\L+\nu_r)} \r(t)
- \b_r \r(t) \i(t) +\Phi_r \r(t)(\nu_i \i(t)+\nu_r \r(t)),\nonumber\\
\s(t)&+\i(t)+\r(t)=1.
\end{align}

\BEN
\im The model \eqref{SIRscD} with $\Phi_s=\Phi_i=\Phi_r=1$  will  be  called
scaled model (SM).

\im  The model \eqref{SIRscD} with $\Phi_s=\Phi_i=\Phi_r=0$  will  be  called
first approximation (FA) in the specific case $\Lambda=\mu$.  The FA is thus:
\begin{align}
\label{SIRp}
\s'(t) &=
\L -\b  \s(t)\i(t) + \g_r \r(t)- {(\g_s +  \L)}\s(t), \nonumber\\
\i'(t) &=
 \i(t)\pp{\b \s(t) +\b_r \r(t) - \pr{\g  +{\nu_i  +\L}}}, \nonumber\\
\r'(t) &= \g   \i(t) +\g_s \s(t) -{(\g_r+\L+\nu_r)} \r(t)
- \b_r \r(t) \i(t),\nonumber\\
\s(t)&+\i(t)+\r(t)=1.
\end{align}

\im  The model \eqref{SIRscD} with $\Phi_s=\Phi_r=1$ and $\Phi_i=0$  will  be
called  intermediate approximation (IA).

\EEN
\eeD

\beR Each model, in particular SIR/V+S,  has a  SM, FA and IA version, which will be denoted by  SIR/V+S-SM, SIR/V+S-FA,  SIR/V+S-IA. \eeR

\beR The   FA is not a constant population model when $\nu_c >0$, for some compartment $c$.  \eeR

\beR \la{r:const} It follows from $N'(t)/N(t)=(\L  -\mu)-\nu_i  \i(t)-\nu_r \r(t)$ that $N'(t)=0$ for all
$t \in \realnneg$ if and only if $\mu=\L $ and $\nu_i=\nu_r=0$.
Thus, the popular assumption of constant population
size~\cite{hethcote1976qualitative}  applies only to epidemics
without extra-deaths, which contradicts  the essence of most
epidemics \cite[Ch. 10.2]{Chavez}. Clearly, constant population papers have in mind   some large $N$ or short
term approximation, but this is rather vague.
\Oth   the FA approximation \eqr{SIRp} with $\mu=\L$, as well as IA, may be
heuristically justified  as   approximations  obtained by ignoring  certain
quadratic terms in  \eqr{SIRsc}. This justifies   studying  FA
without restrictive assumptions like $\nu_i=0$.
\eeR
\beR
 A considerable part of the epidemics literature studies \eqr{SIRNe} with $N(t)=1$ (this produces an analog of \eqr{SIRp} with $\mu \neq \L$).  The justification for studying this model is of course an assumption that $N$ is ``approximately constant". The  purpose of our paper is not to assume that;
  however, we  chose to include results about them, under the name  ``classic/pedagogic models" (PM), to be  in line with this part of the literature. As explained, we need here only  results on the FA model (which approximate the object of interest to us \eqr{SIRsc}),
 and these may be easily recovered  by replacing $\mu$ with $\L$.
\eeR

%\blue{GB: I don't understand this comment: FOA also has $\nu_i=\nu_r=0$? R: no!!}

We conclude this section  by illustrating in Figure~\ref{f:PIP} a comparison
between the trajectories of the first approximation, of the
intermediate approximation, and  of the scaled model.  Note that the
approximate dynamics are an accurate approximation of the SM at  the
beginning of the epidemic (i.e., when $\i(t) \approx 0$). This period
starts with the lower  part in Figure \ref{f:PIP}, and continues
until  the processes start turning towards their distinct endemic
points -- see \cite{Kuehn} for a rigorous slow-fast analysis of
similar models.\fn[4]{\Oth in real life controlled epidemics, \fe\ via  ICU constraints (see \cite{Miclo,AFG}, etc), one has, at least for the French situation with 400 positives out of 100.000 individuals (which is still 4 to 10 times the admissible figures for Japan or other countries), that $i_{\max}\approx 0.004$. So, one may argue that if state upper-constraints are imposed, $i\approx 0$ for all time, not just the beginning.}
%PIP.nb
 \figu{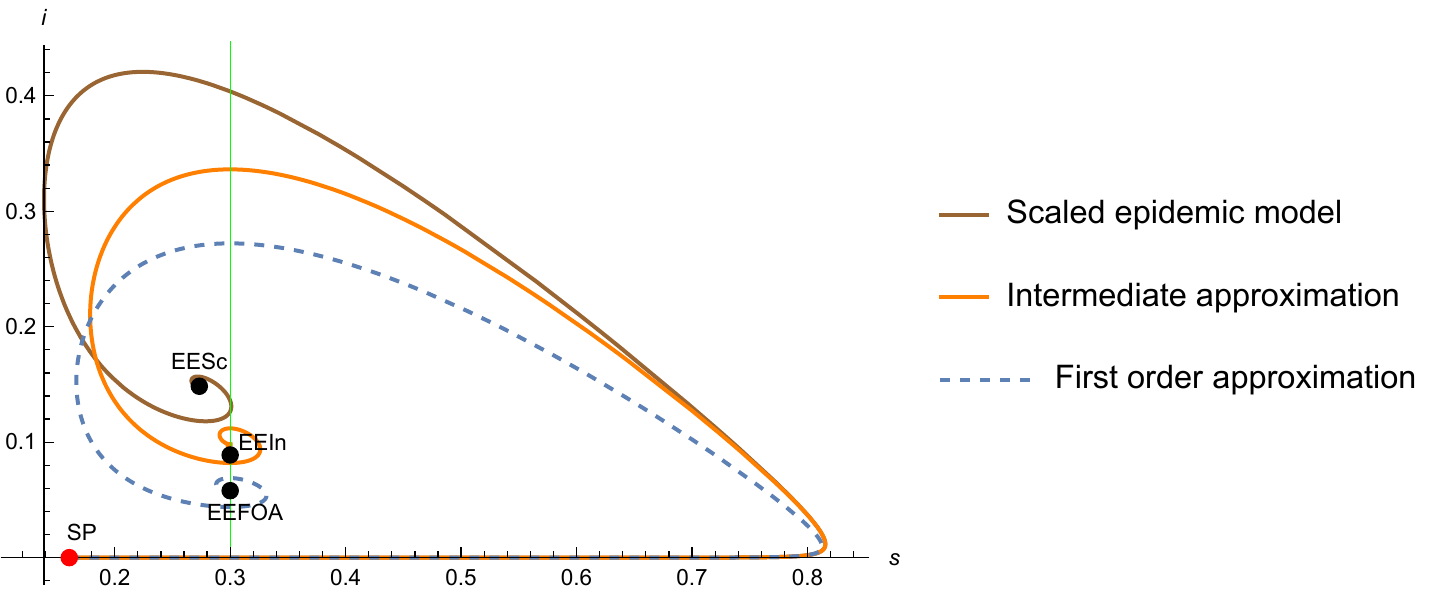}{Parametric $(\s,\i)$ plots of the scaled epidemic \eqr{SIRsc} and   its FA \eqr{SIRp} and intermediate approximations for a SIR-type model with one infectious class, starting from a starting point SP with $i_0=10^{-6}, \mR=3.21,$ critical vaccination  $0.622222$, and $\g_s=.01$. The  other parameters are
 $ \b=5,\; \g  = 1/2,\; \L=\mu=1/10, \; \g_r=1/6, \; \nu_i=.9, \; \nu_r=0$. $EESc, EEIn, EEF0A$ are the stable endemic points of the scaled model, intermediate model, and the FA model, respectively.
 The green vertical line denotes the immunity threshold $1/\nR= s_{EEF0A}= s_{EEIn}.$
 Note that    the epidemic will  spend at first a long time (since births and deaths have  slow rates as compared to the disease) in  the vicinity of the manifold $\vi(t) =0$, where the three processes are indistinguishable, before turning towards the  endemic  equilibrium point(s).}{1}

%\section{The equilibria of the SIR/V+S scaled  model and their stability \la{s:Razee}}\label{subsec2}

\iffalse
%SIRS+VSC3D.nb
\bea %J^{(SC)}=
The Jacobian matrix of the  scaled model \eqr{SIRsc} is
\left(
\begin{array}{ccc}
 i \left(\nu _i-\beta \right)-\Lambda +r \nu _r-\g_s & s \left(\nu _i-\beta \right) & \g_r+s \nu _r \\
 \beta  i & -\g+2 i \nu _i-\nu _i-\Lambda +r (\nu _r+\b_r)+\beta  s & i (\nu _r +\b_r)\\
 \g_s & \g+r (\nu _i -\b_r)& i (\nu _i-\b_r)-\Lambda -\g_r+ \nu _r(2 r-1) \\
\end{array}
\right).
\eea
\fi

\subsection{The disease free system and its equilibria \la{s:Razdfe}}

It is  convenient to eliminate $\r$ from $\s+\i+\r=1$,  and work with the following two-dimensional scaled dynamic
\be{SIRsc2}
\bc
\s'(t)= \L  - \s(t) \i(t)(\b-\nu_i) -(\g_s+\L) \s(t) +( \g_r+\nu_r \s(t)) (1-\s(t)-\i(t))
\\
\i'(t)= \i(t)\pp{ \s(t)\b+\nu_i \i(t)+(\nu_r +\b_r)(1-\s(t)-\i(t))}-\i(t) \pr{\L+\g+\nu_i}
\ec,
\ee
defined on  the positively invariant region \cite{Raz}
\bea
\mD=\left\{(s,i),\; s\geq 0,\; i\geq 0,\; s+i \leq 1 \right\}.
\eea

The fixed points are the solutions of
\be{SIRsc3}
\bc
  \s \pp{ \nu_r  \s  +\i(\b-\nu_i+\nu_r)+\L+\g_s+\g_r-\nu_r }+\i \g_r-(\L +\g_r)=0
\\
 \i\pp{ \s(\b-\nu_r-\b_r) +(\nu_i-\nu_r-\b_r)\i+ \nu_r +\b_r-(\L+\g+\nu_i)}=0
\ec.
\ee

The disease free system ( {with $\i=0,\r=1-\s$}) reduces to
\be{SIRDF}
\begin{split}
\s'(t)= \L   -(\g_s+\L) \s(t) +( \g_r+\nu_r \s(t)) (1-\s(t))\\
= \L  + \g_r -(\g_s+\g_r +\L-\nu_r) \s(t) - \nu_r \s(t)^2,
\end{split}
\ee
and its fixed points  are such that  $\s$ \sats\  the  equation
$$\bc \L +\g_r- \s \pp{ \nu_r  \s  +\L+\g_r+\g_s-\nu_r}=0& \nu_r >0\\
\s \pp{\L+\g_r+\g_s}-(\L +\g_r)=0& \nu_r =0\ec.$$

 One root
\be{sdfe}
 s_{dfe}= \bc \fr{\L +\g_r}{\L+\g_r+\g_s}& \nu_r =0\\ \frac
{\sqrt{\D_{dfe}} -
   \pr{\Lambda +\g_r+\g_s-\nu _r}}
   {2 \nu _r}, \qu \D_{dfe}=4 \nu _r \left(\Lambda +\g_r\right)+
\left(\Lambda +\g_r+\gamma
   _s-\nu _r\right)^2&\nu_r > 0 \ec
\ee
is always  in $[0,1]$ and will be denoted  by $s_{dfe}$.  %SIRS+V-RazR.nb

\beR $s_{dfe}$ is continuous in $\nu_r$, since for $\nu_r$ small,
$$s_{dfe} \approx \frac
{\Lambda +\g_r+\gamma
   _s-\nu _r+\fr{2 \nu _r \left(\Lambda +\g_r\right)}{\Lambda +\g_r+\gamma
   _s-\nu _r} -
   \pr{\Lambda +\g_r+\g_s-\nu _r}}
   {2 \nu _r} \to \fr{\Lambda +\g_r}{\Lambda +\g_r+\gamma
   _s}$$
\eeR

\beR The other root in the  quadratic case $\nu_r>0$
\be{dfe2}
  \frac
{\nu _r- \pr{\Lambda +\g_r+\g_s}-\sqrt{4 \nu _r \left(\Lambda +\g_r\right)+
\left(\Lambda +\g_r+\gamma
   _s-\nu _r\right)^2}
  }
   {2 \nu _r}
\ee
is strictly negative, unless
\be{bop}\bc \L+\g_r=0\\ \nu_r \geq \g_s + \L+\g_r\ec \Eq \bc \L=\g_r=0\\  \nu_r \geq \g_s\ec,\ee in which case it yields  a second DFE point with $\s=0=\i$.\eeR

 \beA \label{r:exc} We will exclude from now on  the particular boundary case \eqr{bop}, which may be resolved by elementary explicit eigenvalues computations -- see \cite{Raz}\fn[4]{Note however that while not necessarily interesting  from an epidemics point of view, this case is remarkable however   mathematically.(\Mp the extra DFE point $(0,0)$ may be either  source or saddle-point, and  there are  two endemic points, which  may be  either a sink and a saddle, or two sinks  \cite{Raz};
   finally, for the general SIR/V+S model,    both cases may be achieved  as small perturbations of the particular case  \eqr{bop})}.
   \eeA

  Once this case is removed, the DFE  is unique, and we may apply the \ngm\ method, the first step of which consists in checking the local asymptotic stability of  $s_{dfe}$  for the {\em   disease-free equation} \eqr{SIRDF}.

In our case, this amounts to proving that
\be{lneg} \l_-:=\nu_r -(\L+\g_r +\g_s) -2 \nu_r s_{dfe}  <0. \ee
 This is automatic both when
$\nu_r=0$, and when $\nu_r>0, $ since by \eqr{sdfe}
$ \l_-=-\sqrt{\D_{dfe}}<0$.
\iffalse
\be{ngmcond} \bc \gamma _s\leq \Lambda + \g_r \quad \text{ (small ``vaccination" rate), or}\\
\gamma _s>\Lambda + \g_r, \qu
 \nu _r<\frac{(\Lambda +\gamma _r+\gamma _s)^2}{\gamma _s-\Lambda -\gamma _r} \ec. \ee

 \beR {Thus, stability of DFE  is not guaranteed if the dying  rate $\nu_r$ of the vaccinated/recovered  is huge and if the ``vaccination rate" $\g_s> \frac{(\Lambda +\gamma _r+\gamma _s)^2}{\gamma _s-\Lambda -\gamma _r}$ is very big. Granted, this is hopefully unrealistic, but not impossible, if the vaccin is totally wrong and actually increases the rate of death ($\nu_r >> \nu_i$). So, we are just trying here to explore theoretically the limits of our mathematical model.} \eeR
\fi

\section{Local DFE stability via the \ngm\ approach \cite{Van}\la{s:ngm}}

Recall (cf. Assumption \ref{r:exc}) that we exclude the case  $\L=\g_r=0,  \nu_r \geq \g_s$ \eqr{bop}, so that
 the  DFE defined in  \eqr{sdfe} is unique and locally stable in the disease free space \eqr{lneg}, so that the \ngm\ approach may be attempted.

 This proceeds in the following steps:
\BEN
 \im  Separate the ``infected equation" for $\i$ \eqref{SIRsc} %and on \eqr{R1}
 as a difference of two terms, $\mathcal{F}, \mathcal{V}$,    as follows
\bea \i'(t) && = \i(t)\pp{ \s(t)\b+(\nu_r +\b_r)(1-\s(t)-\i(t))}-\i(t) \pp{\L+\g+\nu_i-\nu_i \i(t)}\\&& =\mathcal{F}(s,i)-
\mathcal{V}(i)\eea where $
\bc
\mathcal{F}(s,i):= \i(t) \pp{ \s(t)\b+(\nu_r +\b_r)(1-\s(t)-\i(t)) } \\
\mathcal{V}(i):= \i(t) \pp{\L+\g+\nu_i +\nu_i \i(t) }\ec.
$
The necessary decomposition conditions  on $\mF,\mV$ \eqr{cond}
 are immediate,
and the gradients
 $$F=\left[\frac{\partial \mathcal{F}}{\partial \i}\right]_{i=0,s=s_{dfe}}=\b s_{dfe}+(\nu_r+\b_r) \rd,%+2\nu_i \i,
 V=\left[\frac{\partial \mathcal{V}}{\partial \i}\right]_{i=0,s=s_{dfe}}=\g   +\nu_i  + \L, $$
 \saty  $F\ge 0, V > 0$, as required.

\EEN

One may conclude  then, by
the NGM result  \cite[Thm 1]{Van08}, that:
\beP   The DFE
 is  locally stable if  the Perron-Frobenius eigenvalue of the \ngm\ \sats
   \be{mRSIRS} \m0:=\l_{P}(F V^{-1})= \fr{\b s_{dfe} +(\nu_r+\b_r) \rd}{\g   +\nu_i  + \L }%=\fr{\nu_r+\b_r+\sd(\beta-\nu_r-\b_r)}{\Lambda+\ga+\nu_i}
   < 1, \ee
   and is unstable if $\m0 >1$.%\fn[3]{For including the critical case $\m0=1,$ one also needs checking that the perturbation from the linearization \sats\{$f(s,i)=i(F-V) - \mF +\mV=-\nu_i i^2 \leq 0,$ which  only holds here when $\nu_i=0$.}}

\eeP

\beR \la{r:gs}   The equality  $  \m0 = 1$ is linear in $\g_s$ and yields %Gene-case.nb
\be{crvsc} \g_s = \frac{\left(\L+\nu_i+\g-\b \right) \left(\nu _r \left(\gamma +\nu _i-\gamma _r-\nu _r\right)-\beta _r \left(\Lambda +\gamma _r+\nu _r\right)+\beta  \left(\Lambda +\gamma _r\right)\right)}{\left(\beta -\beta _r-\nu _r\right) \left(-\gamma -\nu _i-\Lambda +\beta _r+\nu _r\right)}:=\g_s^*,%:=\fr{f_1 f_2}{f_3 f_4}
\ee
provided that the denominator doesn't blow-up. The parameter $\g_s^*$ is called
 ``critical vaccination".
 %{W.

 The daunting formula \eqr{crvsc} simplifies and turns out  to provide crucial help for  stability analysis   in  the following particular cases\fn[3]{This should  be true in general, but we have not been able to work  analytically with this daunting formula.}
\be{crv} \bc\nu_r=0&\Lra  \fr{\g_s^*}{\Lambda +\gamma _r}=
\frac{\beta-\pr{\Lgn} }{\Lgn-\beta _r}\\\b_r=0&\Lra \fr{\g_s^*}{\Lambda +\gamma _r}=\nR-1, \nR=\frac{\beta }{\Lgn}\ec. \ee
The first formula is positive if either $\b_r \le \Lgn \le \b$, or $\b \le \Lgn \le \b_r$, and  a geometric consequence of this is provided in Section \ref{s:nr0}.
\iffalse
The last formula coincides with the critical vaccination of the FA model discussed in section \ref{s:FOA}.\fn[5]{In that case, it may also be checked directly that the characteristic polynomial and therefore the $\m0$ and critical vaccination  are the same as for FA (this is because the only term different is $J^{SC}_{1,2}$, and this does not change the characteristic polynomial).}
\fi
\iffalse
Finally, we note that we may also define an analog critical immunity parameter (as a function of
$\g_s$), which will \saty:
\be{crl} \bc\nu_r=0&\Lra  \fr{\g_s}{\Lambda +\gamma _r^*}=
\frac{\beta-\pr{\Lgn} }{\Lgn-\beta _r}\\\nu_r=\b_r=0&\Lra \fr{\g_s}{\Lambda +\gamma _r^*}=\nR-1\ec.
\ee
\fi
 \eeR

\beR[Direct  stability analysis for the DFE] \la{s:R0sc}
The Jacobian of \eqr{SIRsc2} (with $\r$ eliminated) is

{\small
\be{Jsi}
J= \left(
\begin{array}{cc}
 -i \left(\beta -\nu _i+\nu _r\right)-\gamma _s-\Lambda -\gamma _r+\nu _r-2 s \nu _r & -s \left(\beta -\nu _i+\nu _r\right)-\gamma _r \\
 i \left(\beta -\beta _r-\nu _r\right) & s \left(\beta -\beta _r-\nu _r\right)-\Lambda-\gamma +(2 i-1) \left(\nu _i-\nu _r-\b_r\right) \\
\end{array}
\right).
\ee
}
  Plugging $i=0$  yields
$$ J_{dfe}=\left(
\begin{array}{cc}
-2 s \nu _r +\nu _r-\Lambda -\g_r-\g_s &
   -s \left(\beta -\nu _i+\nu _r\right)-\g_r
   \\
 0 & s \left(\beta
   -\nu _r-\b_r \right)-\g-\L-\nu _i +\nu _r+\b_r  \\
\end{array}
\right),$$
with eigenvalues
 $ \bc \l_P=s_{dfe} \left(\beta -\nu _r-\b_r\right) -\g-\nu _i-\Lambda +\nu _r+\b_r \\
 =s_{dfe} \beta +(\nu _r+\b_r) \rd -\g-\nu _i-\Lambda\\\l_-=-2 s_{dfe} \nu _r +\nu_r -\Lambda -\g_r -\g_s  \ec.$

We recover  the same result as via the \ngm\ method. The DFE is locally stable iff $\l_- <0$, and
\be{mR1}
\l_P<0 \Leftrightarrow \fr{\sd  \beta +(\nu _r+\b_r)\rd}{\Lgn} <1.
\ee
\eeR
\begin{figure}[H] %SIRS+VS2.nb
    \centering
        \includegraphics[scale=1]{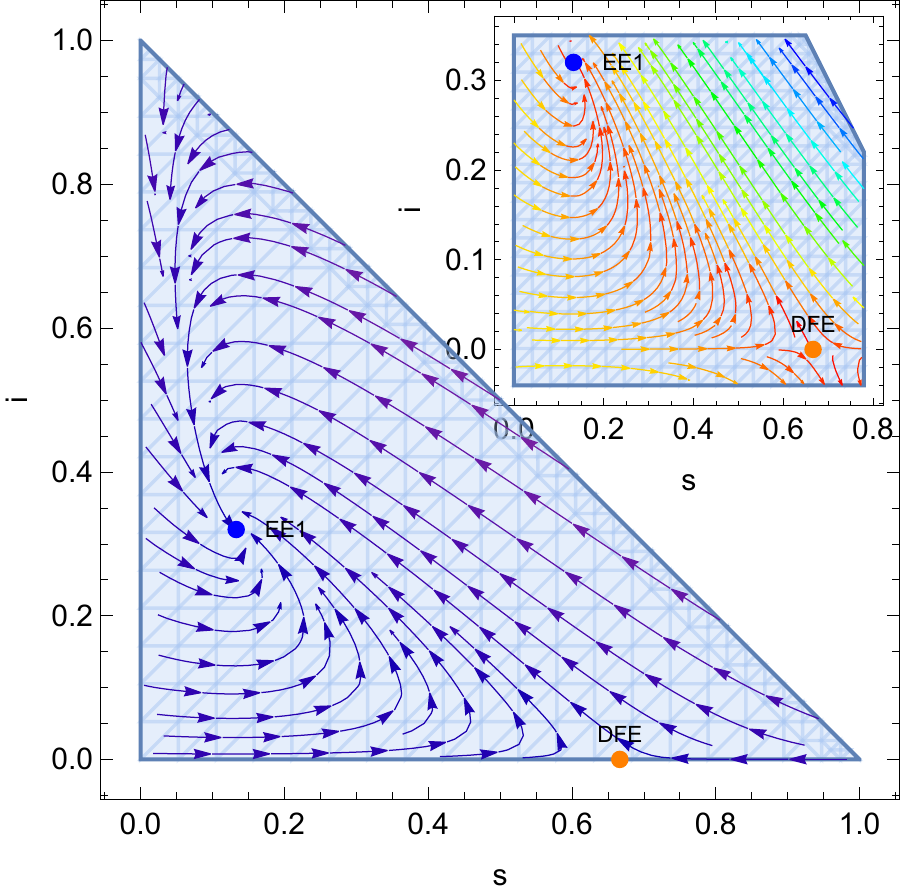}
        \caption{Stream plots of $(s,i)$ for the scaled model \eqr{SIRsc2}
   illustrating the case when $\g_s $ is less than the critical vaccination $\g_s^*$ defined in \eqr{crvsc}, EE1 is a  stable equilibrium point, DFE is a (boundary) saddle point, and EE2 is outside the  domain.}
        \label{fig:spgc}
    \end{figure}

\iffalse
{As an aside, we could work alternatively eliminating $s$. This yields  the Jacobian
$$
J= \left(
\begin{array}{cc}
 -\gamma +2 i \nu _i-\nu _i-(\beta  (2 i+r-1))-\Lambda +r \left(\beta _r+\nu _r\right) &
  i \left(-\beta +\beta _r+\nu _r\right) \\
 \gamma +r \nu _i-\gamma _s & i \nu _i-\Lambda -\beta _r-\gamma _r+2 r \nu _r-\nu _r-\gamma _s \\
\end{array}
\right),$$ and
$$ J_{dfe}=\left(
\begin{array}{cc}
s\pr{\beta -\beta _r-\nu _r} -\gamma -\nu _i  -\Lambda +\beta _r+\nu _r & 0 \\
 \gamma +r \nu _i-\gamma _s & 2 s \nu _r+\nu _r-\Lambda -\beta _r-\gamma _r-\gamma _s \\
\end{array}
\right),$$
with eigenvalues
 $  \l_P,\red{\l_+=2 s \nu _r+\nu _r-\Lambda -\beta _r-\gamma _r-\gamma _s},$}
{and after more work the same result is obtained}.
\fi
\subsection{Global stability of the \DFE}

\beP \label{p:nop}  If $\mathcal{R}_0<1$ and DFE is the unique equilibrium point, then it is globally asymptotically stable in the invariant set $\mathcal{D}$.
\eeP

\Prf Each solution starting in $\mathcal{D}$ is obviously bounded so its $\omega$-limit set is not empty. The Poincare-Bendixon Theorem implies that this is the unique equilibrium point DFE, since else  it would be a closed orbit ( see \cite{HS}) and one may check similarly as in  \cite[Thm 3.1]{Der} that there exist no periodic orbits.
\QED

Since the explicit conditions for the  uniqueness of the equilibrium point are pretty complicated, we prefer to give them  only in a particular case, in  section \ref{s:nr0}.

For the general case, we may find a simple criteria if we weaken the concept of global stability of the \DFE\ as follows
\beP \la{p:sgs} If $$\max[ \beta _r+ \nu_r, \b,\Lambda+\gamma+\nu_i]=\Lambda+\gamma+\nu_i,$$ then the DFE is ``strongly globally  stable",  in the sense that the  function $L(\s,\i)=\i$  is Lyapunov
     over the invariant region  $\mD:=\left\{(\s,\i)\in \mathbb{R}_+^2: \; 0\leq \s+\i\leq 1\right\}.$
 \eeP
 \beR In lay terms, we may say that  ``the epidemics never picks up" in this case. \eeR

\Prf The non-negative function $L(\s,\i)=\i$, with $L(1,0)=0,$  is Lyapunov (i.e. the epidemics may never increase) iff
\bea
L'(\s,\i) =\i' \le 0 \Eq s (\beta -\beta _r-\nu_r) +(\nu_i-\nu_r -\b_r)  i + \beta _r+\nu_r-\Lambda-\gamma-\nu_i \le 0.
\eea

The  maximimization of $\i'$ reduces thus to a simple linear programming problem. Thus, there   must exist an extremal point of   $\mD$ where the maximum   is attained, and
\bea &&\max_{(s,i) \in \mD} s (\beta -\beta _r-\nu_r) +(\nu_i-\nu_r -\b_r)  i + \beta _r+\nu_r-\Lambda-\gamma-\nu_i=\\&&
\max[ \beta _r+\nu_r-\Lambda-\gamma-\nu_i, \beta -\Lambda-\gamma-\nu_i, -\Lambda-\gamma] \leq 0,\eea
provided that
\be{LIg} \beta_r+ \nu_r \leq \Lambda+\gamma+\nu_i,\m0<1 \Eq \max[ \beta _r+\nu_r, \b,\Lambda+\gamma+\nu_i]=\Lambda+\gamma+\nu_i.\ee \QED

\section{The endemic equilibria \la{s:end}}

The endemic equilibrium set may be obtained algebraically by solving one of the variables from 
$\i'/\i= 0$  in \eqr{SIRsc3}, and plugging into the other equation.  Eliminating $\s$ using
$$s=1-i+\frac{i (\nu _i-\b)+\beta -\pr{\Lgn} }{-\beta +\beta _r+\nu _r}, \b \neq \b_r+\nu_r,$$
yields a quadratic equation   %DerRiF.nb
 $A i^2 +B i +C=0 $
 where $ A=\left(\beta -\beta _r\right) \left(\beta -\nu _i\right) \left(\beta _r+\nu _r-\nu _i\right),$
 and the other coefficients are very complicated. 
   
   Thus, in the complex plane, $i^{(EE)}_{1,2}=\frac{-B \pm \sqrt{\D}}{2 A}.$
   Since locating analytically the endemic points  requires considerable effort, we will restrict starting with the next section to the case $\nu_r=0$.
   
   \iffalse 
   Eliminating $\i$ by
$$\i=1+ \frac{\L+\g+s(\nu_r+\b_r-\b)}{\nu_i-\nu_r-\b_r}=1-s+ \frac{\L+\g+s(\nu_i-\b)}{\nu_i-\nu_r-\b_r}, \; \nu_i\neq \nu_r+\b_r$$
yields the quadratic equation  in $s$ %SIRSsc.nb
 $A s^2 +B s +C=0, $
 where \be{ABC}\bc
 C=-\Lambda  \left(\beta _r+\gamma _r+\nu _r-\nu _i\right)-\gamma  \gamma _r \\ B=\nu _i^2+\beta  \left(-\gamma -\nu _i-\Lambda +\beta _r+\gamma _r\right)-\nu _i \left(-\gamma +\beta _r+\gamma _r+\gamma _s\right)+\beta _r \left(\Lambda +\gamma _s\right)%\\-\nu _i \gamma _r-\beta  \left(\gamma +\Lambda -\gamma _r\right)+(\gamma +\Lambda ) \beta _r+\pr{\beta -\gamma -\nu _i+\gamma _s}\pr{-\nu _i+\beta _r+\nu _r}
 \\ A=\left(\beta -\beta _r\right) \left(\beta -\nu _i\right)
   \ec.\ee
   \fi
   
   \beR  We note however already that: 
   \BEN  \im The equation is quadratic   iff  $\beta \neq \beta _r, \beta \neq \nu _i $ and $\beta _r+\nu _r\neq \nu _i$. We will exclude at first these particular cases,   but note  that  they suggest that  different regimes  occur when the respective thresholds are crossed -- this will be further explained below.
   \im  Locating the endemic points may be achieved geometrically by studying   the intersection of  the vertical and horizontal isoclines:

  \begin{enumerate}%[(a)]
  \im
  the  $i'=0$ isocline is given by $i=0$ and by the herd immunity line

\be{iiso} s \left(\beta -\beta _r-\nu_r\right) +i (\nu_i-\beta _r-\nu_r)  =\Lambda+\gamma+\nu_i-\beta _r-\nu_r,\ee
with slope $\fr{\beta -\beta _r-\nu_r}{\beta _r+\nu_r-\nu_i}$.
\item  the $s'=0$ isocline is
\be{hyp} \fr{\nu_r}{\L+\g_r+\g_s} s^2 + i s \fr{\beta -\nu _i+\nu_r}{\L+\g_r+\g_s}+s +i \fr{\g_r}{\L+\g_r+\g_s}-\fr{\L+\g_r}{\L+\g_r+\g_s}=0,\ee
which is a hyperbola.

 This passes by definition  through the DFE, and intersects therefore necessarily the domain
 when $$ \sd <1 \Eq \g_s >0.$$

\end{enumerate}
\EEN

   \eeR

\section{The  case $\nu_r=0, \g_s>0$\la{s:nr0}}
%derNur0.nb

%Recall that the DFE is locally stable/unstable when  the basic reproduction number  $\m0=  \fr{\b s_{dfe} +\b_r \rd}{\g   +\nu_i  + \L }$ is smaller/larger than $1$.

Getting sharper stability results beyond the weak $\m0$ alternative requires locating
  the  endemic points, and this seems quite  difficult in general. Therefore, we will restrict \fno\ to the case $\nu_r=0$, which avoids the necessity
  of handling the complications arising from the square root formula of $\sd$. The case $\g_s=0$,
   essentially covered in \cite[Sec. 4]{Der}, requires special treatment -- see Section \ref{s:SIRI}.

 The  quadratic equation   %DerRiF.nb
 $A i^2 +B i +C=0 $
 has coefficients \be{ABCi} \bc C=-\left(\gamma +\nu _i+\Lambda \right) \left(\Lambda +\gamma _r+\gamma _s\right)+\beta  \left(\Lambda +\gamma _r\right)+\beta _r \gamma _s\\B=\beta  \left(-\gamma -\nu _i-\Lambda +\beta _r-\gamma _r\right)-\beta _r \left(\nu _i+\Lambda +\gamma _s\right)+\nu _i \left(\gamma +\nu _i+2 \Lambda +\gamma _r+\gamma _s\right)\\A=\left(\beta -\nu _i\right) \left(\nu _i-\beta _r\right)\ec.\ee

The endemic points are still complicated:
  \bea
  EE_{1,2}= \bc  s= \frac{\beta  \left(\gamma +\nu _i+\Lambda -\beta _r-\gamma _r\right)-\nu _i \left(\gamma +\nu _i-\beta _r-\gamma _r-\gamma _s\right)-\beta _r \left(\Lambda +\gamma _s\right)\pm \sqrt{\D}}{2 \left(\beta -\beta _r\right) \left(\beta -\nu _i\right)}  \\
  i= \frac{\beta  \gamma +\beta  \Lambda +\beta  \nu _i-\gamma  \nu _i-2 \Lambda  \nu _i-\nu _i^2+\nu _i \beta _r+\Lambda  \beta _r-\beta  \beta _r\mp\sqrt{\D}}{2 \left(\nu _i-\beta \right) \left(\beta _r-\nu _i\right)} \ec,
  \eea
where
{\small \beq \la{eesc}
 &&\D=4 \left(\beta -\beta _r\right) \left(\beta -\nu _i\right) \left(\Lambda  \left(-\nu _i+\beta _r+\nu _r\right)+(\gamma +\mu ) \gamma _r\right)\\&& +\left(\beta  \left(-\gamma -\nu _i-\mu +\beta _r+\gamma _r+\nu _r\right) +\beta _r \left(-\nu _i+\mu +\gamma _s\right) +\nu _i \left(\gamma +\nu _i-\gamma _r-\gamma _s\right)-\nu _r \left(\gamma +\nu _i-\gamma _s\right)\right){}^2.\no\eeq}

 By solving $C=0$, we may compute however an  important critical value for $\b$,
\be{bc} \beta_* =\frac{\left(\gamma +\nu _i+\Lambda \right) \left(\Lambda +\gamma _r+\gamma _s\right)-\beta _r \gamma _s}{\Lambda +\gamma _r}%= \b_r +\frac{\left(\Lambda +\gamma _r+\gamma _s\right) \left(\Lgn -\beta _r\right)}{\Lambda +\gamma _r},
\ee
above which one (the higher) of the endemic points crosses above the $i=0$ axis, entering therefore $\mD$ (equivalently, $\m0$ become larger than $1$).

Locating the endemic points may be attempted algebraically --see Lemma \ref{l:alg} in the Appendix seemed quite difficult.

 We resorted therefore   to a geometric study of the isoclines in Theorem \ref{cases}, attempting to
 explain geometrically all the possible cases. For example, the case $\m0>1$ turned out equivalent to the unicity of the endemic point,  and to the fact that the immunity line intersection with the domain lie on both sides of the $s'=0$ isocline.
 \beR \la{r:odd} Since the number of crossing points must be odd on one hand, and less than two on the other, identifying such a crossing is equivalent to the uniqueness of the endemic   point.
 \eeR
 We turn now to listing some elementary geometric facts, in particular the coordinates of various intersections  points.
\BEN \im
 When $\nu_i\neq \b$, the $s'=0$ isocline becomes the hyperbola
   \bea && \sd= i s \fr{\beta -\nu _i}{\L+\g_r+\g_s} + \fr{\g_r}{\L+\g_r+\g_s}i +s\\ &&
   =i  \fr{\beta -\nu _i}{\L+\g_r+\g_s}\pr{s + \fr{\g_r}{\beta -\nu _i}} +s+ \fr{\g_r}{\beta -\nu _i}- \fr{\g_r}{\beta -\nu _i}
   \\&& \Eq
    \pr{i  \fr{\beta -\nu _i}{\L+\g_r+\g_s}+1}\pr{s + \fr{\g_r}{\beta -\nu _i}} =\sd+\fr{\g_r}{\beta -\nu _i},\eea
   with asymptotes $s=\fr{\g_r}{\nu_i -\b}, i=\fr{\L+\g_r+\g_s}{\nu_i -\b}$. Note that the center is in the first quadrant when $\b < \nu_i$ and in the third quadrant \oth, and that the intersection $E$ with the line $s=0$ has $i_E=1+ \fr\L {\g_r}$, outside $\mD$.
   Also important will be the convexity of the branch which intersects $\mD$. From $$i''(\sd)=2\fr{\sd+k}{(s+k)^3} \fr {\L+\g_r+\g_s}{\b-\nu_i}, k= \fr {\g_r}{\b-\nu_i}, $$ we find that our branch is convex when $\b >\nu_i$ and concave  \oth.
   The  equality case $\nu_i= \b$  is analyzed in the following remark.

 \beR \la{r:bn}  When $\nu_i= \b$,  the $s'=0$ hyperbola degenerates into a line {$s(\L+\g_r)+ i \g_r-\g_r-\L=0 $.}  The intersection of the two lines gives a unique endemic point $EE= \left(\frac{(\gamma +\Lambda ) \gamma _r}{\Lambda  \left(\b_r-\nu _i\right)}+1,\frac{(\gamma +\Lambda ) \left(\Lambda +\gamma _r\right)}{\Lambda  \left(\nu _i-\b_r\right)}\right)$ {which never belongs to the feasible region}. Indeed $i_{ee} >0 $ if and only if $ \nu_i > \b_r $, and this implies further
 $$s_{ee}+i_{ee}=1+  \frac{(\gamma +\Lambda)\L }{\L(\nu_i-\b_r)}>1 \Lra EE \notin \mD.$$

 \iffalse
\BEN

\im  (this   ensures also $s_{ee} <1$). % which we will assume \fno\ in the rest of the proof;

%\im $s_{ee} >0$ requires further $$\g \g_r< \red{\L(\nu_i-\b_r-\g_r)},$$  and thus \wmh\ $\nu_i>\b_r+\g_r$;
\EEN
 the union of the three cases is disjoint since  $\L \nu_i+(\g+\L)(\g_r+\L)<\b_r\L +(\g+\L)(\g_r+\L)<\L \nu_i$ is impossible and $\left\{\b_r>\nu_i\right\}\cup \left\{\b_r<\nu_i\right\}=\emptyset$, then $EE$ cannot belong to the domain.
 \fi
 \eeR

\im A crucial role in the analysis is played by the point  where the immunity line intersects $i=0$, given by $B( \frac{\L+\nu_i+\g-\b_r}{\b-\b_r},0), \b_r \neq \b $ (when $\b_r =\b $, $B$ goes to $\I$). This point \sats:

\be{bB}\bc s_B<0 &
\text{if } \L+\nu_i+\g < \b_r <\b, \text{ or } \b <  \b_r <  \L+\nu_i+\g\\
s_B \in [0,1]&\text{if } \b \leq  \L+\nu_i+\g <\b_r, \text{ or } \b_r <  \L+\nu_i+\g < \b\\s_B >1 & \text{if } \b_r < \b <\L+\nu_i+\g \text{ or } \L+\nu_i+\g<\b < \b_r \ec.
\ee

The six cases listed above are the basis of our geometric  analysis provided in Theorem \ref{cases}.
\beR \la{r:sBd}  $s_B$ coincides with $\sd$ iff $\g_s=\g_s^*$, which fits with the fact that $\g_s^* \in (0,1)$ iff one of these two cases occurs -- recall Remark \ref{r:gs}.
 \eeR

\im  Another important point is the point $A$ where  the immunity line \eqr{iiso} intersects $i=1-s$, with coordinates
\be{lpA} (s_A,i_A)=\pr{\frac{\gamma +\Lambda }{\beta -\nu _i},\frac{\beta -\pr{\gamma
  +\nu _i+\Lambda} }{\beta -\nu _i}}.\ee
  \Ity  $A \in \mD$ iff $\b>\L+\nu_i+\g$, that it moves then to the fourth  quadrant when $\L+\nu_i+\g\ge \b>\nu_i$, and that it jumps from the fourth to the second quadrant when  $\b$ decreases below the threshold $\nu_i$.

\im The point where the immunity line intersects $s=0$ is $C(0,1+\frac{\gamma+\Lambda }{\nu _i-\beta _r}).$ This is negative when $\nu_i < \b_r < \L+ \g+\nu_i$, in the domain when $\b_r \ge \L+ \g+\nu_i$,   and bigger than $1$ when $\b_r \le \nu_i$. \Mp
$$C=\fr{\L + \g + \nu_i -\b_r}{ \nu_i -\b_r}\text{  \sats\ }\bc i_C\le 0 &
\text{if }  \nu_i <  \b_r <  \L+\nu_i+\g\\
i_C \in [0,1]&\text{if } \b_r \geq  \L+\nu_i+\g\\i_C >1 & \text{if } \b_r < \nu_i\ec.$$

\im   The unique point $D$ where the hyperbola intersects $\s+\i=1$ within the domain has coordinates \be{lpD} \bc s_D=\fr12+\frac{\Lambda +\gamma _s-\sqrt{4 \Lambda  \left(\nu _i-\beta \right)+\left(\beta -\nu _i+\Lambda +\gamma _s\right){}^2}}{2 \left(\beta -\nu _i\right)}\\
i_D=\fr12+\frac{-\Lambda -\gamma _s+\sqrt{4 \Lambda  \left(\nu _i-\beta \right)+\left(\beta -\nu _i+\Lambda +\gamma _s\right){}^2}}{2 \left(\beta -\nu _i\right)}\ec.\ee

\EEN

\beR \la{r:bb} When $\beta =\beta _r$, the slope  $\fr{\beta -\beta _r}{\beta _r-\nu_i}$  of the immunity line becomes $0$
and  $i_C=i_A=\frac{\beta -\pr{\gamma
  +\nu _i+\Lambda} }{\beta -\nu _i}=\fr 1{\gamma
  +\nu _i+\Lambda} \frac{\m0 -1 }{\beta -\nu _i}.$
  The dynamical system admits a unique endemic point and a unique EE given by
\bea  EE=\left(  \frac{\Lambda  \left(\beta -\nu _i+\gamma _r\right)+\gamma  \gamma _r}{\left(\beta -\nu _i\right) \left(\beta -\gamma -\nu _i+\gamma _r+\gamma _s\right)},\frac{\beta-\Lgn}{\beta -\nu _i} \right),
\eea  %bbr.nb
which may be checked  to belong to the feasible region iff $$\nR=\frac{\b}{\Lgn}>1.$$ Indeed, $i_{ee} >0$ requires either $\b < \nu_i$, which leads subsequently to a contradiction, or $\b >\Lgn$, which may be shown to imply $s>0, s+i <1$.
\iffalse
Similarly, $s>0$ can arise in two ways, one of which is impossible, and the other yielding
$\gamma <\frac{\Lambda  \gamma _s}{\Lambda +\gamma _r}, \nu _i\leq \frac{\Lambda  \gamma _r+\gamma  \gamma _r}{\Lambda }$.
\fi
The stability analysis of this case is elementary and left to the reader.
\eeR

The formula \eqr{bB} and the subsequent Remark \ref{r:sBd} suggest splitting the analysis according to the order of the three quantities $\b, \Lgn, \b_r$ and on the orders of $\g_s, \g_s^*=\left(\Lambda +\gamma _r\right) \frac{\beta-\pr{\Lgn} }{\Lgn -\b_r}$ and of $(\nu_i,\b)$. We end up with ten cases, nine of which are fully resolved
 in Theorem \ref{cases}, and one of which is left partly open.
Note that, as proved in the Appendix, these ten cases
form a disjoint decomposition of the parameter space, if only strict inequalities are allowed.

Before stating Theorem \ref{cases}, we provide graphical illustrations of the 10 cases.

\begin{figure}[H] %DerRi.nb
    \centering
    \begin{subfigure}[a]{0.30\textwidth}
        \includegraphics[width=\textwidth]{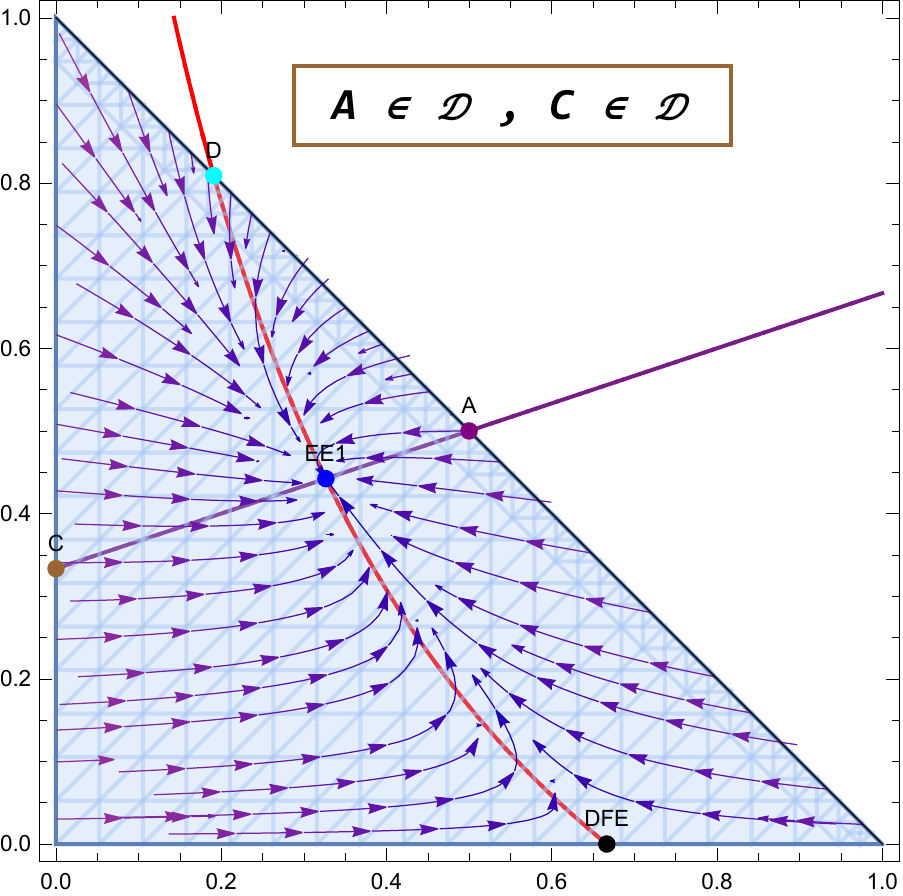}
        \caption{$\b=5 >\b_r=4 >3=\L+ \g+ \nu_i=1+1+1, \g_r=1,    \g_s=1>\g_s^*\Eq  s_B>1, \b>\b_r>\g+\nu_i+\L, $.}
        \label{fig:cas1}
    \end{subfigure}
     ~
    \begin{subfigure}[a]{0.30\textwidth}
        \includegraphics[width=\textwidth]{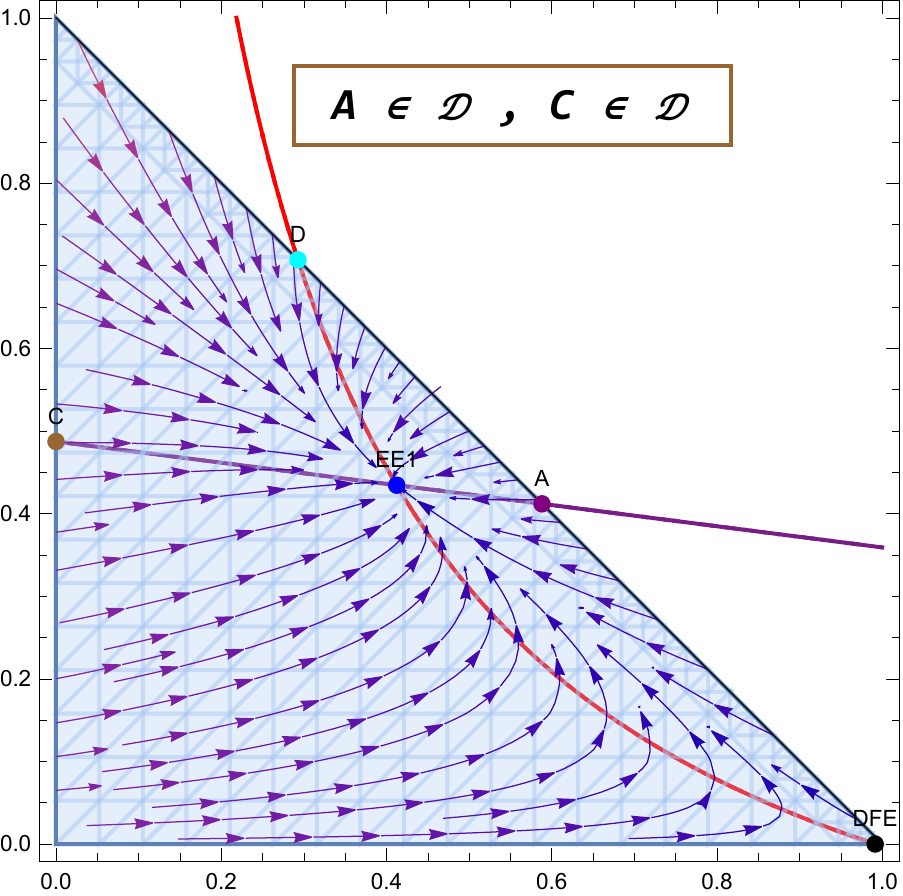}
        \caption{$\b_r=4 >\b=3.5>0.21=\L+ \g+ \nu_i=1+1+0.1,  \g_r=1/6, \g_s= 0.01.$  }
        \label{fig:cas2}
    \end{subfigure}
    
    \begin{subfigure}[b]{0.30\textwidth}
        \includegraphics[width=\textwidth]{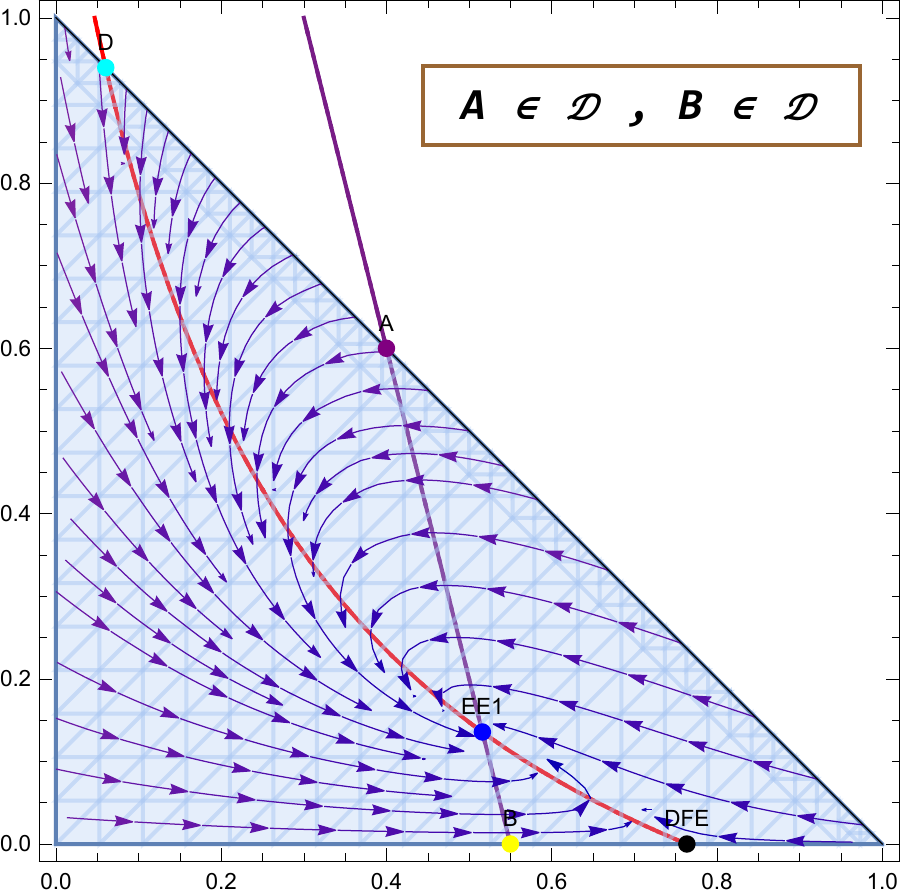}
        \caption{$B,A \in \mD$, with $\b=0.3 > 0.21 =\L+ \g+ \nu_i=0.01+0.05+0.15 >\b_r=0.1,  \g_r=1/26, \g_s.=.015 <\g_s^* \Eq s_B<s_{dfe}$.% The DFE is a saddle point, and EE1=$(0.51611,0.13556)$ is a stable spiral point.
        }
        \label{fig:cas3}
    \end{subfigure}
     ~
    \begin{subfigure}[b]{0.29\textwidth}
        \includegraphics[width=\textwidth]{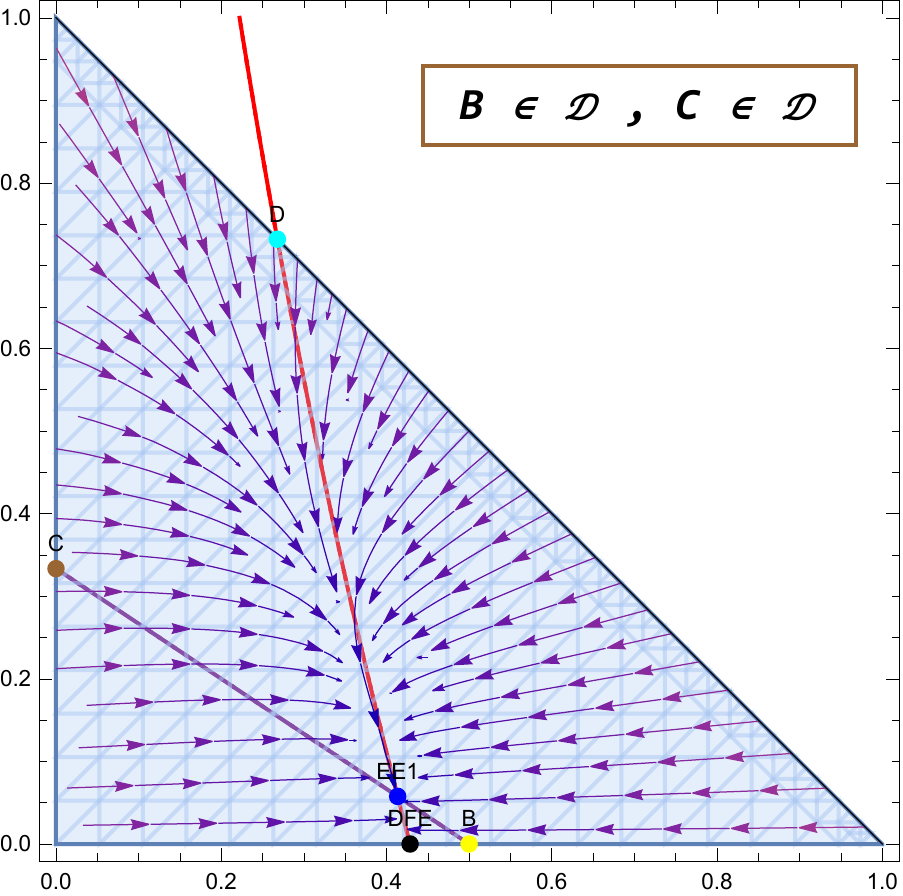}
        \caption{$B,C \in \mD$, with $\m0=1.04762>1, \b=2< \L+ \g+ \nu_i=1+1+1=3 < \b_r=4, \g_r=1/2,   \g_s= 2>\g_s^* \Eq s_B>s_{dfe}$.}
        \label{fig:casBR}
    \end{subfigure}
     \caption{Stream plots of $(\s,\i)$ when $ \g_s>0$, in  four cases when $\m0>1$. The DFE is a saddle point and EE is a stable sink.}
     \label{fig:3}
    \end{figure}

%\usepackage{tabularx}
\iffalse   %DerRi.nb
\begin{tabularx}{0.8\textwidth} {
  | >{\raggedright\arraybackslash}X
  | >{\centering\arraybackslash}X
  | >{\raggedleft\arraybackslash}X | }
 \hline
  & $\nR<1$ & $\nR>1$ \\
 \hline
 $\b>\b_r,\; \b>\nu_i$  & DFE is the only global stable node  & DFE is a saddle point, and there is a stable endemic sink  \\
\hline
$\b<\b_r,\; \b<\nu_i$  & DFE is the only global stable sink & unfeasible  \\
\hline
$\b>\b_r,\; \b<\nu_i$  &  DFE is the only global stable sink  & unfeasible  \\
\hline
$\b<\b_r,\; \b>\nu_i$  & DFE is a sink, one endemic point which is a sink, and one endemic point which is a saddle point with \blue{$s_B<s_{dfe}$} & DFE is a saddle point, and one endemic point which is a sink with \blue{$s_B>s_{dfe}$} \\
\hline
\end{tabularx}
\fi
%\figu{cas4L}{StreamPlot of $(s,i)$ when  $\b<\b_r,\; \b>\nu_i$, $\nR<1$ and $s_B<s_{dfe}$.}{.6}
%\figu{cas4R}{StreamPlot of $(s,i)$ when  $\b<\b_r,\; \b>\nu_i$, $\nR>1$ and $s_B>s_{dfe}$.}{.6}

%when R0<1
     \begin{figure}[H] %DerRi.nb
    \centering
    \begin{subfigure}[a]{0.23\textwidth}
        \includegraphics[width=\textwidth]{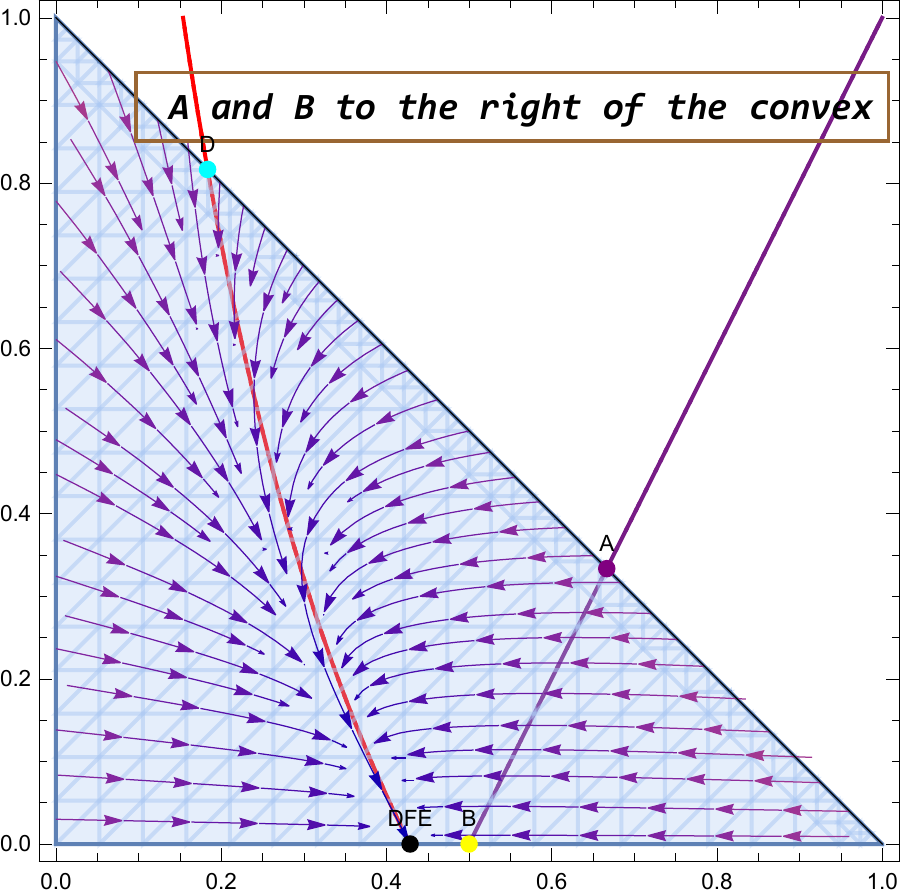}
        \caption{$\m0=0.952381<1,\b=4, \b_r=2, \L+\g+\nu_i=1+1+1=3, \b_r^{(+)}=1.625, \g_r=1/2, \g_s= 2, \b_r<\L+ \nu_i+\g <\b, \nu_i<\b, \b_r> \b_r^{(+)}, \;\text{and}\;  \g_s>\g_s^*\Eq s_B>s_{dfe}$.}
        \label{fig:cas2BR}
    \end{subfigure}
    ~
    \begin{subfigure}[a]{0.23\textwidth}
        \includegraphics[width=\textwidth]{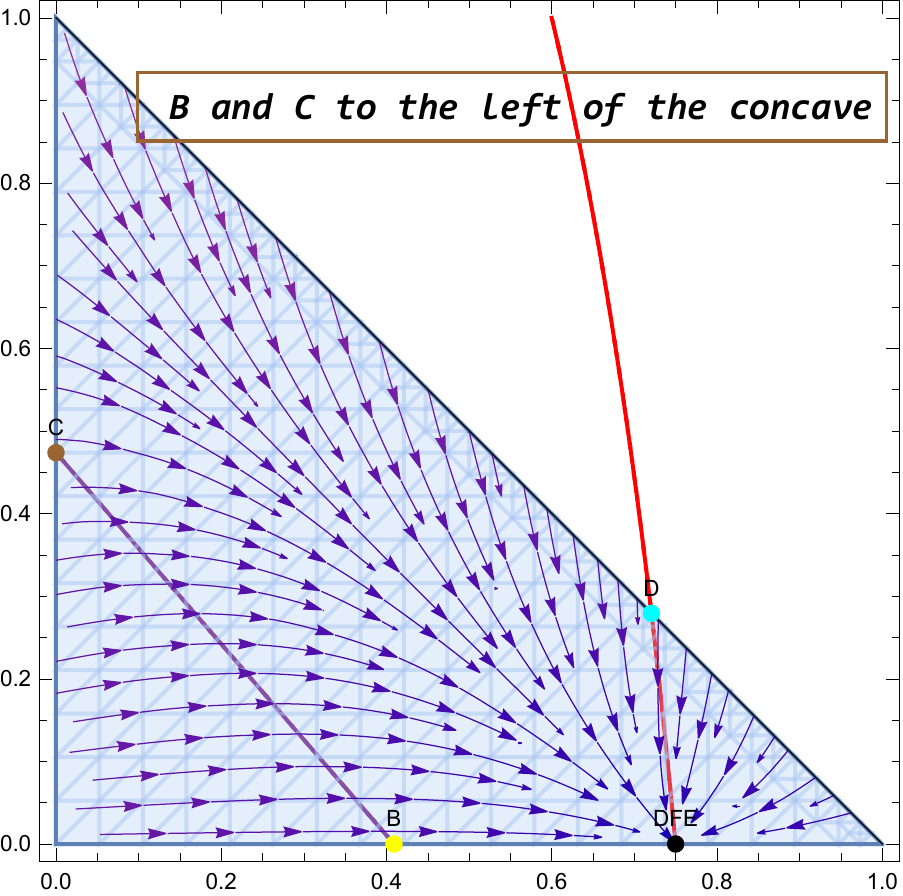}
        \caption{$\b=0.03, \b_r=0.25, \L+\g+\nu_i=0.03+0.07+0.06=0.16, \b_r^{(+)}=0.0219597, \g_r=0.03, \g_s= 0,02, \b<\nu_i <\L+ \nu_i+\g < \b_r, \b_r> \b_r^{(+)}, \;\text{and}\;  \g_s<\g_s^* \Eq s_B<s_{dfe}$.}
        \label{fig:casNointCo}
         \end{subfigure}
         ~
      \begin{subfigure}[a]{0.23\textwidth}
        \includegraphics[width=\textwidth]{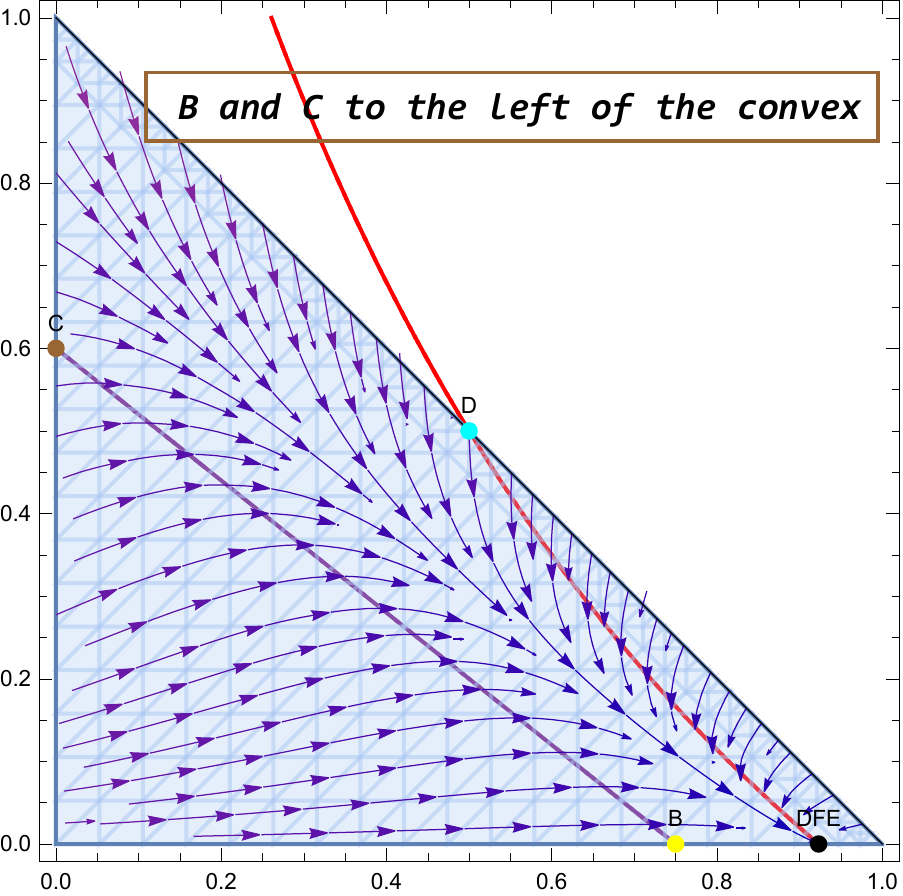}
        \caption{$\b=0.1, \b_r=0.3, \L+\g+\nu_i=0.03+0.07+0.05=0.15, \b_r^{(+)}=0.729883, \g_r=0.03, \g_s= 0,02, \nu_i<\b <\L+ \nu_i+\g \leq \b_r, \b_r< \b_r^{(+)}, \;\text{and}\;  \g_s<\g_s^* \Eq s_B<s_{dfe}$.}
        \label{fig:casNoint}
         \end{subfigure}
         
         \begin{subfigure}[b]{0.23\textwidth}
        \includegraphics[width=\textwidth]{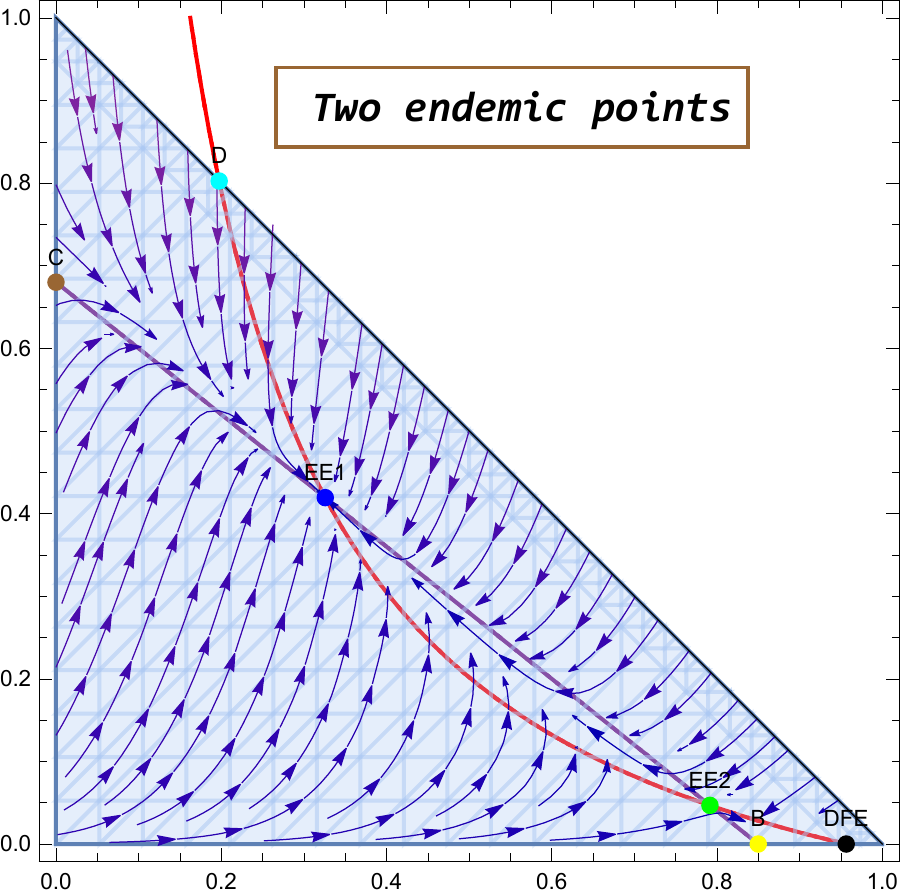}
        \caption{$\b=0.2, \b_r=0.4, \L+\g+\nu_i=0.01+0.07+0.15=0.23, \b_r^{(+)}=0.327443, \g_r=10^{-4}, \g_s= 0,002,  \nu_i<\b<\L+\g+\nu_i<\b_r, \b_r\geq \b_r^{(+)}, \; \text{and}\;  s_B<s_{dfe}$.}
        \label{fig:2endpts}
    \end{subfigure}
     ~
     \begin{subfigure}[b]{0.23\textwidth}
        \includegraphics[width=\textwidth]{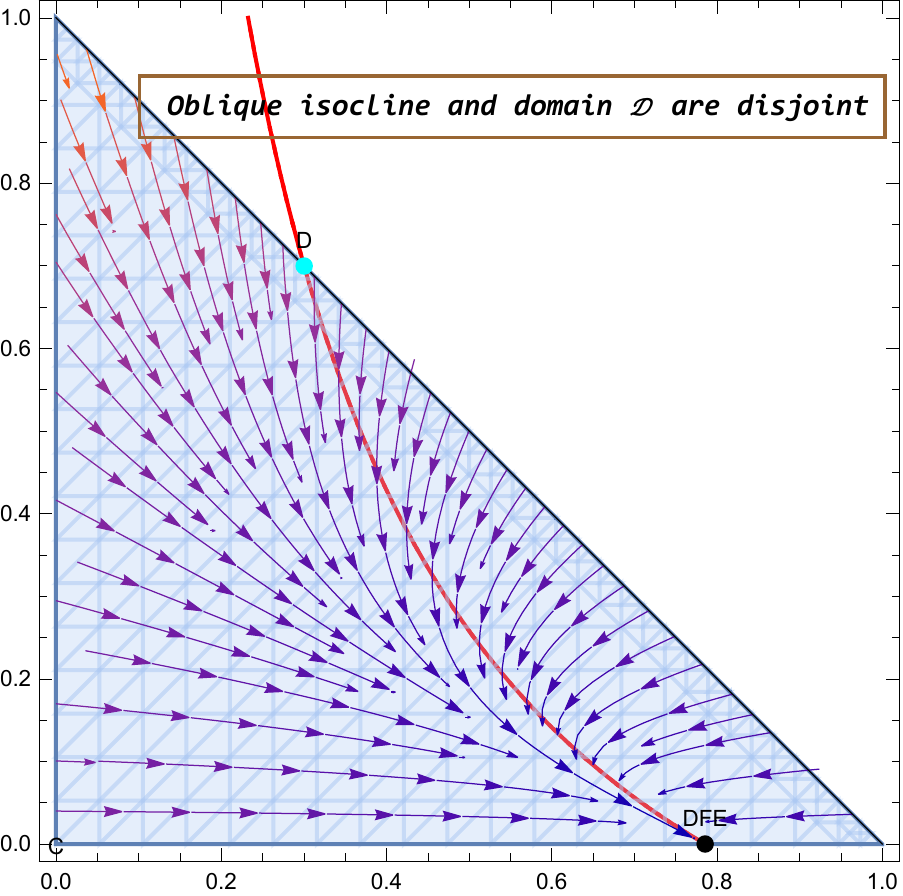}
        \caption{$\L+\g+\nu_i=1+2.1+0.1=3.2>\b_r=3.1>\b=3,  \g_r=0.1, \g_s=0.3,  \b<\b_r<\L+\g+\nu_i, \; \text{and}\;   \b>\nu_i$.}
        \label{fig:casconv}
    \end{subfigure}
    ~
    \begin{subfigure}[b]{0.23\textwidth}
        \includegraphics[width=\textwidth]{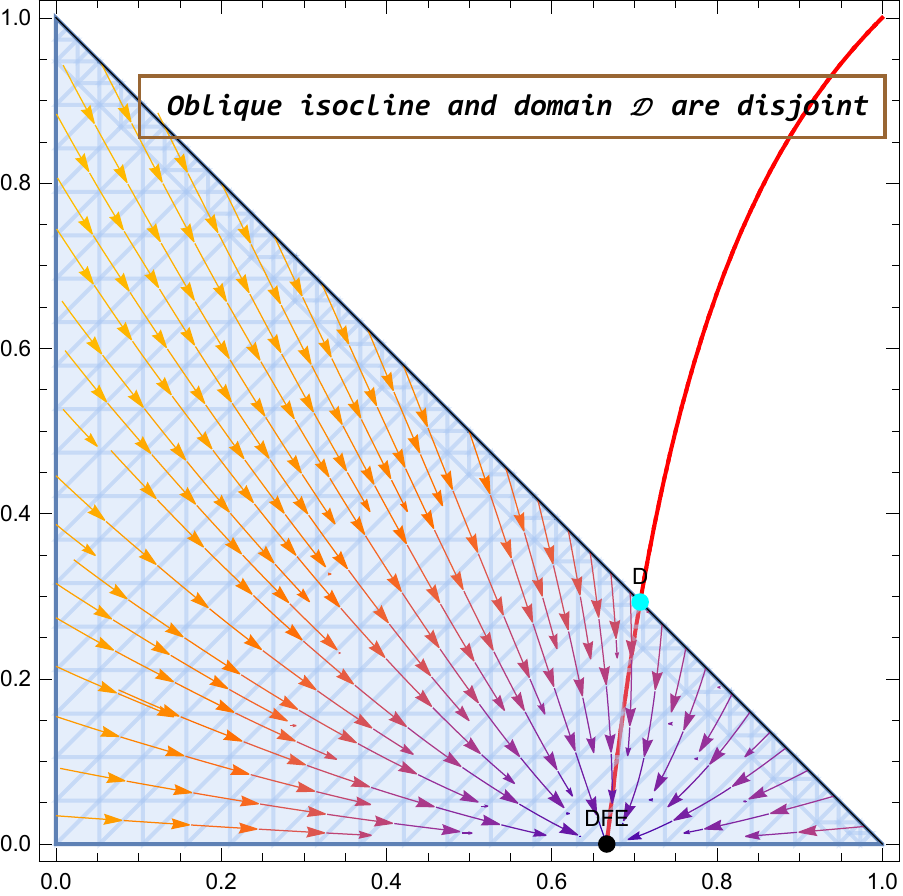}
        \caption{$\L+\g+\nu_i=1+1+2.5=4.5>\b=0.5>\b_r=0.25,  \g_r=1, \g_s=1,  \b_r<\b<\L+\g+\nu_i, \; \text{and}\;   \b<\nu_i$.}
        \label{fig:casconc}
         \end{subfigure}

    \caption{Stream plots of $(\s,\i)$ when $ \g_s>0$, in six cases when $\m0<1$. The DFE is the only stable sink in the boundary of the domain, with the exception of case (d), when  EE1=($0.313823,0.428942$) is a second sink, and EE2=($ 0.799177,0.0406582$) is a saddle point.}
    \label{fig:4}
       \end{figure}

\figu{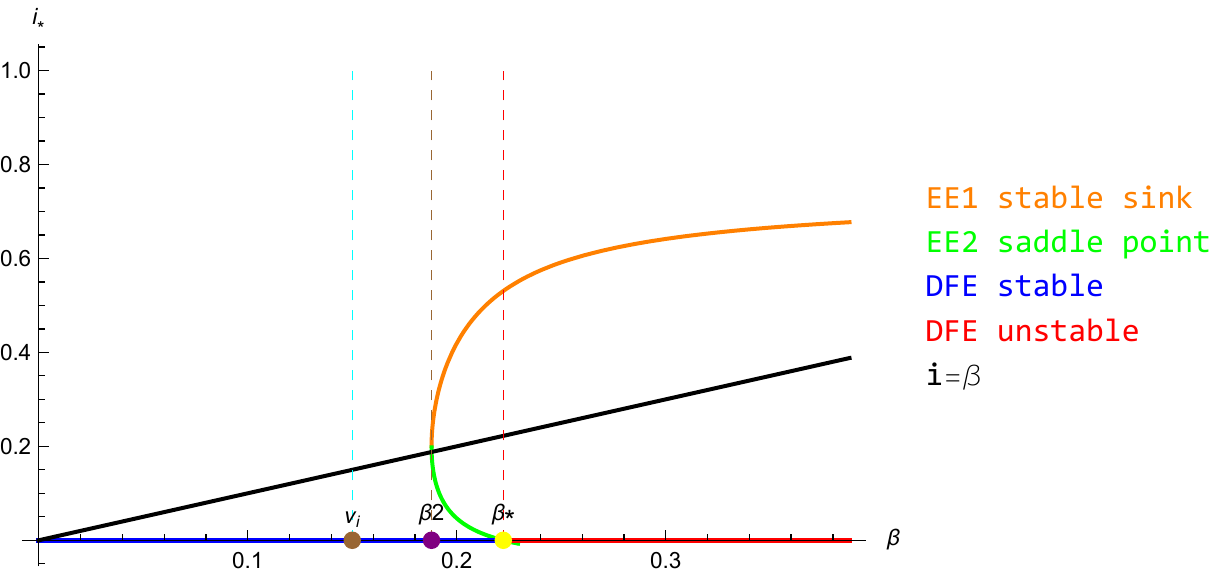}{Diagram bifurcation with respect to $\b$, in the case $\max[\L+\g+\nu_i,\b_r^{(+)}]<\b_r$. For $\b< \nu_i$ we are in the case of Fig. 4(b) (Thm. \ref{cases}.2.b)), with the immunity line to the left of the hyperbola, and no endemic points. The same situation occurs for $\nu_i <\b <\b_2$, where $\b_2$ is the largest root of $\Delta(\b)=0$, except that the hyperbola changes to convex -- see Fig. 4(c) (Thm. \ref{cases}.2.c)(i)).  After $\b_2$ two endemic points emerge -- see Fig. 4(d) (Thm. \ref{cases}.2.c)(ii)).
 The lower endemic point exits through the boundary $i=0$ at $\b *$, defined in \eqr{bc}, after which the remaining endemic point remains the only stable point.}{1} %DerJust1.nb 
\beT \la{cases}
Suppose $\nu_r=0$, and that neither two of the three parameters $\beta, \beta _r, \nu _i$ coincide. Then, one of the following  cases must arise:
  \BEN \im $\m0>1 \Eq $  precisely one   endemic point lies in $\mD$, which may occur in one of the following four   ways:
  \BEN \im   $AC \in \mD$ crosses the hyperbola
   with positive slope $ \L+\g+\nu_i\le \b_r< \b$ --see Figure 3(a);%1

  \im $AC \in \mD$ crosses the hyperbola with negative slope $\L+\g+\nu_i\le \b< \b_r$ --see Figure 3(b);%2

  \im    $\b_r \le \L+\g+\nu_i \le \b, \g_s <\g_s^*\Eq $ $BA \in \mD$ crosses the hyperbola -- Fig. 3(c);%3

   \im  $ \b \le  \L+\g+\nu_i \le \b_r, \g_s \ge \g_s^* \Eq $ $BC \in \mD$ crosses the hyperbola -- Fig. 3(d).%6
\EEN

In all these cases, the  endemic point is a sink and DFE is a saddle point.

  \im
  $\m0\leq 1$ is the union of the following six cases (similar to the above, but taking also into account the convexity of the hyperbola branch, in some cases):
  \BEN
\im  $$ \b_r < \Lgn<\b, \nu_i<\b,$$
when both the points $A,B$ lie to the right of a convex hyperbola branch, and   the isoclines do not intersect in $\mD$--see Figure 4(a).

\im  $$\b<\nu_i < \Lgn< \b_r,$$
when both the points $C,B$ lie to the left of a concave hyperbola branch, and   the isoclines do not intersect in $\mD$--see Figure 4(b).
\im When both the points $C,B$ lie to the left of a convex hyperbola branch, we have two subcases:
\BEN \im  When
 $$
 \nu_i< \b  < \Lgn<\b_r, \g_s<\g_s^*, \D<0, $$
 the isoclines  intersect  $\mD$, but  do not intersect each other
 --see Figure 4(c).

\im The isoclines   intersect in $\mD$, yielding two endemic points --see Figure 4(d). Necessary conditions for this are
 $$
 \nu_i< \b { < \Lgn<\b_r}, \g_s<\g_s^*, \D\ge 0, $$
 and {we conjecture that the necessary and sufficient conditions   are obtained by adding
$ \b_r  \ge \b_r^{(+)} $,
where $\b_r^{(+)}$ is defined in \eqr{brp}}.

 In this case, the DFE is one of two sink points, whose attraction domains are separated by the separatrices of the third  fixed saddle point.
\EEN

\im

The $i'/i=0$  isocline does not intersect the interior of $\mD$ in the following two cases:

\BEN
\im
$ \beta \le  \beta _r\le  \Lgn$, with hyperbole  concave  -- see figure 5(a)-- or convex, according to whether $\b > \nu_i$ or not;
\im  $\beta_r <  \beta \le \Lgn$, with hyperbole  convex -- see figure 5(b)-- or convave, according to whether $\b < \nu_i$ or not;

\EEN

\EEN

\im In all the cases when the
   DFE is the unique fixed point within the domain, it is a globally stable sink.

 \EEN

 \eeT

     \Prf
  \BEN \im
As noted already,  $\b>\L +\g + \nu_i$ is equivalent to fact that the point $A=(\frac{\gamma +\Lambda }{\beta -\nu _i},\frac{\beta -\nu _i-\gamma -\Lambda }{\beta -\nu _i})$  lies in  $\mD$, which applies in the cases 1.(a), 1.(b), 1.(c). \Fr
\wmc in this case the point $A$
 is always to the right (above) the hyperbola, i.e.
 \bea s_D=\fr12+\frac{\Lambda +\gamma _s-\sqrt{4 \Lambda  \left(\nu _i-\beta \right)+\left(\beta -\nu _i+\Lambda +\gamma _s\right){}^2}}{2 \left(\beta -\nu _i\right)}<\frac{\gamma +\Lambda }{\beta -\nu _i}=s_A.\eea

To conclude the existence of a unique endemic point, it is enough then to find in these three cases a point of the immunity line  below the hyperbola.
Referring to Figures 3, we see that the following cases may arise:
 \BEN

 \im $AC\in \mD$ crosses the hyperbola, and $s_B < 0$.  We must then be in the case $\Lgn <\b_r <\b$, which implies $\b_r \geq  \L+\nu_i+\g$, and so  $C \in \mD$ is below the hyperbola  --see Figure 3(a).

  \im  $AC\in \mD$ crosses the hyperbola, and $s_B \geq 1$. We must then be in the case $\Lgn <\b <\b_r$, which implies again $\b_r \geq  \L+\nu_i+\g$, and so  $C \in \mD$ is again below the hyperbola -- see {Figure 3(b)}.

  \im  $BA\in \mD$ crosses the hyperbola when $\m0>1, \b > \Lgn > \b_r, s_B \leq \sd \Eq  \g_s \leq \g_s^*$, since $B$ is below the hyperbola --see Figure 3(c).
\im  $BC\in \mD$ crosses the hyperbola (and $A \notin \mD$) when $\m0>1, \b_r > \Lgn > \b, s_B > \sd \Eq  \g_s \geq \g_s^*$, since $B$ is below the hyperbola --see Figure 3(d).
    \EEN

 \im  We turn now to the  case $ \m0 \leq 1$, when the point $A$ lies in the fourth quadrant if $\nu_i < \b$, and in the second quadrant otherwise.
   \BEN

\im In this case $s_B> \sd, s_A > s_D$.  Since the immunity line  is to the right of a convex  hyperbola branch,
    it is clear geometrically that they cannot intersect  within the domain -- see Figure 4(a).

    \im Similarly,  the end points in $\mD$ of the immunity line  are to the left of a concave  hyperbola branch, and so they cannot intersect  within the domain -- see Figure 4(b).

   \im For two endemic points to exist, {it is necessary that}
      $s_B= \frac{\L+\nu_i+\g-\b_r}{\b-\b_r} \in (0,\sd) $, which requires that
      $$ \b \leq \Lgn \leq \b_r, $$
      and this implies that the other intercept $C=(0,\fr{\L + \g + \nu_i -\b_r}{ \nu_i -\b_r})$ is also in $\mD$.

      The $i'/i=0$ isocline intersects then $\mD$, and may also intersect the hyperbola, when
 the discriminant \beq \la{disc} && \D=\left(\nu_i(\nu_i-\b_r+\gamma -\g_r-\g_s)+\b_r (\g_s+\Lambda )+\beta  (\b_r-\gamma +\g_r-\Lambda -\nu_i)\right)^2\no\\&&+4 (\beta -\b_r)(\beta -\nu_i) (\gamma  \g_r+\Lambda  (\b_r+\g_r-\nu_i))
  \eeq
     \sats\  $\D \ge 0$.

     {The inequality $\D > 0$ is quadratic in $\b_r$} and maybe rewritten as
\beq &&\beta _r+\Lambda +\gamma _r-\nu _i+
\frac{\Lambda  \left(\Lambda +\gamma _r\right)}{\beta -\nu _i-\Lambda }-\frac{\gamma  \left(\beta -\nu _i\right)
\left(\beta -\nu _i+\Lambda +2 \gamma _r\right)}
{\left(\beta -\nu _i-\Lambda \right){}^2}\notin [-\mL,\mL] \no
\\&&
\mL=2 \fr{\sqrt{\gamma \left(\Lambda +\gamma _r\right) \left(\beta -\nu _i\right){}^2   \left(\beta -\nu _i+\gamma _r\right) \left(\Lgn -\beta \right)}}
{\left(\beta -\nu _i-\Lambda +\g_s\right){}^2 +4 \Lambda \g_s} \la{brp}
\eeq
(note that $\mL>0$ when $\nR <1, \nu_i < \beta )$.
% \red{It may be shown by considering the average of the two points $s_1^{(EE)}, s_2^{(EE)}$ that in the first interval the endemic points cannot be both inside the domain} %see also Figure \ref{f:der5},and so necessarily }

\iffalse
\bea  \b_r^{(+)}=\mL+ \frac{\gamma  \left(\beta -\nu _i\right) \left(\beta -\nu _i+\Lambda +2 \gamma _r\right)-\left(\beta -\nu _i-\Lambda \right) \left(\nu _i^2-\nu _i \left(\beta +\gamma _r\right)+\beta  \left(\Lambda +\gamma _r\right)\right)}{\left(-\beta +\nu _i+\Lambda \right){}^2}.\eea
\fi

 We conjecture based on numerical evidence  that the two  endemic points belong to $\mD$ only
when $\b_r$ is larger than the largest root $\b_r^{(+)}$ of $\D=0$, defined implicitly in \eqr{brp}
\footnote{When $\g_s=0$, the largest root is reduced to
\bea  \b_r^{(+)}=\mL+ \frac{\gamma  \left(\beta -\nu _i\right) \left(\beta -\nu _i+\Lambda +2 \gamma _r\right)-\left(\beta -\nu _i-\Lambda \right) \left(\nu _i^2-\nu _i \left(\beta +\gamma _r\right)+\beta  \left(\Lambda +\gamma _r\right)\right)}{\left(-\beta +\nu _i+\Lambda \right){}^2}\eea where $$\mL:= \frac{2 \sqrt{\gamma  \left(\beta -\nu _i\right){}^2 \left(\Lambda +\gamma _r\right) \left(-\beta +\Lgn \right) \left(\beta -\nu _i+\gamma _r\right)}}{\left(-\beta +\nu _i+\Lambda \right){}^2}.$$}

 \im  In the last case we must show that the $i'/i=0$  isocline does not intersect the interior of $\mD$. \Equ  \wms
  in each of
   the  two subcases
$$\bc \b<\b_r \le \Lgn\\ \b_r\le \b \le  \Lgn \ec$$
 none of the points $A,B,C$ belongs to $\mD$.
  {This is a tedious computation, not reproduced here. For a quick check, we offer a Mathematica file  on our website (we  rely mostly on FindInstance with empty output to show that certain cases do not exist, and on the command  Reduce to decompose other cases into subcases)}.

\EEN
\EEN
\QED

\section{The boundary case $\g_s=\nu_r=0$\la{s:SIRI}}
%deric.nb

The  case $\g_s=\nu_r=0$, generalizes on  the particular case $\g_s=\nu_r=0=\g_r$, which was   called SIRI model  in  \cite[Sec. 4]{Der}.

Note first that in this case $\sd=1$. \Itf\ the 4, 5 cases (1(d) and 2(a)) may not arise,and the  case 6 (2(b)) becomes degenerate since $\sd=1$.
It becomes now possible that the hyperbola  does not intersect the domain. This is equivalent to its slope
at the DFE $i'(\sd)=-\fr{\L +\g_r}{\b-\nu_i +\g_r}$ being either positive  or less than  $-1 $, and  further equivalent to $$\b \leq \L+ \nu_i.$$ This further implies $ \mR <1$, and,   together with the absence of endemic points and of periodic solutions, leads to the fact that the DFE is  the global attractor.

After having dealt with this case, which includes also the concave case $\beta >\nu _i$, one may restrict  to the case when  {the hyperbola does intersect the domain}, which is equivalent to its slope at the DFE being such that
$$-1 < i'(\sd) < 0 \Eq \b >\L+ \nu_i$$
(note this   implies  that its center is in the third quadrant).

The proof becomes  simpler than in the previous section.
For example, the point $D$ of intersection
of the hyperbola with $i=1-s$ \sats
$$ s_D=\frac{\Lambda }{\beta -\nu _i}  {\in (0,1)},$$
and hence belongs to $\mD$.

  The  roots of $\D=0$ simplify now to:
\be{brc0} \b_r^{(\pm)}:=\nu _i-\Lambda -\gamma _r-
\frac{\Lambda  \left(\Lambda +\gamma _r\right)}{\beta -\nu _i-\Lambda }
+\frac{\gamma  \left(\beta -\nu _i\right)
\left(\beta -\nu _i+\Lambda +2 \gamma _r\right)}
{\left(\beta -\nu _i-\Lambda \right){}^2}\pm \mL,
%\nu _i-\L +\g+ 2 \frac{ \gamma  \Lambda ^2+ \sqrt{\gamma  \Lambda  \left(\beta -\nu_i\right)^3 \left(\Lgn -\beta\right)}}{\left(-\beta +\nu_i+\L \right){}^2}+\frac{\L  (\L -3 \gamma )}{-\beta+\nu _i+\Lambda }.
   \ee

   \bea
\mL=2 \fr{\sqrt{\gamma \left(\Lambda +\gamma _r\right) \left(\beta -\nu _i\right){}^2   \left(\beta -\nu _i+\gamma _r\right) \left(\Lgn -\beta \right)}}
{\left(-\beta +\nu _i+\Lambda \right){}^2}.
\eea

\section{The classic/pedagogical  SIR/V+S model  \la{s:FOA}}

%We revisit here the classic well-studied first orderapproximation \eqr{SIRp}.
The pedagogical model is defined as follows:
\begin{align}
\label{SIRpm}
\s'(t) &=
\L -\b  \s(t)\i(t) + \g_r \r(t)- {(\g_s +  \mu)}\s(t), \nonumber\\
\i'(t) &=
 \i(t)\pp{\b \s(t) +\b_r \r(t) - \pr{\g  +{\nu_i  +\mu}}}, \nonumber\\
\r'(t) &= \g   \i(t) +\g_s \s(t) -{(\g_r+\mu+\nu_r)} \r(t)
- \b_r \r(t) \i(t).
\end{align}

The  following properties hold:%\fn[5]{In our context, only the case $\mu = \L$ is relevant.}
  \BEN  \iffalse
  \im The second equation of \eqr{SIRpm} implies   the so-called {\bf threshold phenomenon}:
$\i(t)$ grows iff \be{RS} \nR(t) :=\s(t) \fr{\b }{\g   +\mu +\nu_i    }:=\s(t) \nR > 1.\ee

The rate $\nR$ will be called {\bf  reproduction number} (under  stochastic models, this  models the expected number of susceptibles infected by one infectious).

Once  the susceptibles $\s(t)$ reach the {\bf immunity threshold}
\be{IT} \fr 1 {\nR},\ee
the  infectious start declining. To  avoid trivialities,  we will only consider the case $\nR >1$, since otherwise $\i(t)$ decreases always.
\fi
\im
The  region \be{mD}\mD = \{(\s, \i, N) \in  \mathbb{R}_{{+}}^3,
\s + \i \leq N \leq\L/\mu\}\ee
 \mbs\  be  positively invariant
with respect to \eqr{SIRpm}; therefore, this region must include an attractor set
\cite{mena1992dynamic, vargas2009constructions}.

\im {\bf The DFE equilibrium point} of the pedagogic system is obtained plugging $i_{dfe}=0$
 in \eqr{SIRpm}. Solving with respect to $(\s,\r)$  the remaining first and third equation
\bea
\bc
0= %\frac{S'(t)}{N}=
\L    -(\g_s+\mu) \s + \g_r \r,
\\0=  \g_s \s -(\g_r+\mu+ \nu_r) \r,
\ec
\eea
yields %StreamPlots.nb
\be{peddfe}  X^{(DFE)}=   \pr{\frac{\Lambda  \left(\mu +\gamma _r+\nu _r\right)}{\mu  \gamma _r+\left(\mu +\nu _r\right) \left(\mu +\gamma _s\right)},0, \frac{\Lambda  \gamma _s}{\mu  \gamma _r+\left(\mu +\nu _r\right) \left(\mu +\gamma _s\right)}}.\ee

In particular, the DFE for the FA model is  obtained by substituting $\L=\mu$ in \eqr{peddfe}:
\be{FAdfe}  X^{(DFE)}=   \pr{\frac{\Lambda  \left(\Lambda +\gamma _r+\nu _r\right)}{\Lambda  \gamma _r+\left(\Lambda +\nu _r\right) \left(\Lambda +\gamma _s\right)},0,\frac{\Lambda  \gamma _s}{\Lambda  \gamma _r+\left(\Lambda +\nu _r\right) \left(\Lambda +\gamma _s\right)}}.\ee %StreamPlots.nb
Note the relation $\nu_r \g_s =0 \Lra i_{dfe}+ r_{dfe}+ s_{dfe}=1.$

%\red{The stability of the DFE (the so-called $\mR$ alternative), will be discussed below.}

\im {When $\b_r>0$ there may be two endemic equilibrium points %\cite[Sec. 4]{Der}
({we omit their complicated expressions}), but when $\b_r=0$ there is a unique FA EE, }  obtained by plugging  $$s=\fr 1\nR=\fr{\g+\L+\nu_i}\b$$ in the first and third equation, and solving the linear system
\be{eqend}
\bc
\L +\g_r \r(t)-\b \i/\nR - (\g_s+\L)/\nR= 0,\\
\g \i +\g_s/\nR -(\g_r+\L+\nu_r) \r=0
\ec.
\ee
This yields
$X^{(EE)}=$

\be{eep}   \pr{ \frac{  1}{\nR  },
\frac{\g_r (\Lambda -\Lambda  \nR)+(\Lambda +\nu_r) (\g_s+\Lambda -\Lambda  \nR)}{\gamma  \g_r \nR-\beta  (\g_r+\Lambda +\nu_r)},\frac{\gamma  \nR (\g_s+\Lambda -\Lambda  \nR)-\beta  \g_s}{\nR (\gamma  \g_r \nR-\beta  (\g_r+\Lambda +\nu_r))}}\ee
\bea =\pr{ \frac{  1}{\nR  },\frac{\nR s_{dfe}-1}{ z},\frac{\gamma  \nR (\g_s+\Lambda -\Lambda  \nR)-\beta  \g_s}{\nR (\gamma  \g_r \nR-\beta  (\g_r+\Lambda +\nu_r))}},
   \eea
where $z:= \nR \left[   \frac{\g_r(\L+\nu_i)+(\L+\nu_r)(\g+\L+\nu_i)}{\Lambda  \gamma _r+\left(\Lambda +\nu _r\right) \left(\Lambda +\gamma _s\right)} \right]$.

\iffalse
\be{eep}   \pr{ \frac{  1}{\nR  },
\frac{\beta  \Lambda  \left(\Lambda +\gamma _r+\nu _r\right)-\left(\Lgn \right) \left(\Lambda  \gamma _r+\left(\Lambda +\nu _r\right) \left(\Lambda +\gamma _s\right)\right)}
{\beta  \left(\left(\nu _i+\Lambda \right) \left(\Lambda +\gamma _r+\nu _r\right)+\gamma  \left(\Lambda +\nu _r\right)\right)},\frac{\beta  \gamma  \Lambda -\left(\Lgn \right) \left(\gamma  \Lambda -\gamma _s \left(\nu _i+\Lambda \right)\right)}{\beta  \left(\left(\nu _i+\Lambda \right) \left(\Lambda +\gamma _r+\nu _r\right)+\gamma  \left(\Lambda +\nu _r\right)\right)}}\ee
\bea =\pr{ \frac{  1}{\nR  },\frac{\nR s_{dfe}-1}{\b z},\frac{1}{\nR}\frac{\g \L(\nR-1)+\g_s (\nu_i+\L)}{  \left(\left(\nu _i+\Lambda \right) \left(\Lambda +\gamma _r+\nu _r\right)+\gamma  \left(\Lambda +\nu _r\right)\right)}}
   \eea
  % \bea =\pr{ \frac{  1}{\nR  },\frac{\gamma _r \L (1 - \nR)+\left(\Lambda +\nu _r\right) \left(\Lambda   -\nR \L+\gamma _s\right)}{\gamma  \nR \gamma _r-\beta  \left(\Lambda +\gamma _r+\nu _r\right)},\frac{\gamma  \nR \left(\Lambda -\nR \L+\gamma _s\right)-\beta  \gamma _s}{\nR \left(\gamma  \nR \gamma _r-\beta  \left(\Lambda +\gamma _r+\nu _r\right)\right)}} \eea
where $z= \frac{\left(\nu _i+\Lambda \right) \left(\Lambda +\gamma _r+\nu _r\right)+\gamma  \left(\Lambda +\nu _r\right)}{(\g +\nu_i+\L)(\L \g_r +(\L+\nu_r)(\L+\g_s))}$.
\fi
 Note the intriguing simplification of $i^{(EE)}$, which shows that $$i^{(EE)} \geq 0 \Eq 1\le \nR s_{dfe},$$
 and \itm  the endemic point belongs to the domain iff $1\le \nR s_{dfe}$.
 This gives a pre-warning on the role of the parameter %\be{R02}
 $\nR s_{dfe}:=\mR$
 in the stability  of the DFE.

\EEN

\beXa
 In the particular case $\nu_r=\g_r=\g_s=0 \Lra s_{dfe}=1$, we recover the SIR-FA example, for which  the sharp threshold property holds  \cite[(4.1)]{Shuai}, i.e.  the \DFE\ is globally
 stable iff $\mR= \nR \; s_{dfe} \leq 1$.

  The endemic point simplifies to
\bea X^{(EE)}= \pr{ \frac{  1}{\nR  },\frac{\L}{ \b}\pr{\nR-1},\frac{\g}{ \b}\pr{\nR-1}}.\eea
% {\pr{ \frac{  1}{\nR  },\L \frac{\nR s_{dfe}-1}{\b z},\frac{\g (\nR-1)}{\b}}}={\pr{ \frac{  1}{\nR  },\L \frac{\nR -1}{\b z},\frac{\g (\nR-1)}{\b}}}. \eea
\eeXa
We may observe that the endemic point $X^{(EE)}$ is positif if $\nR>1$.

\figu{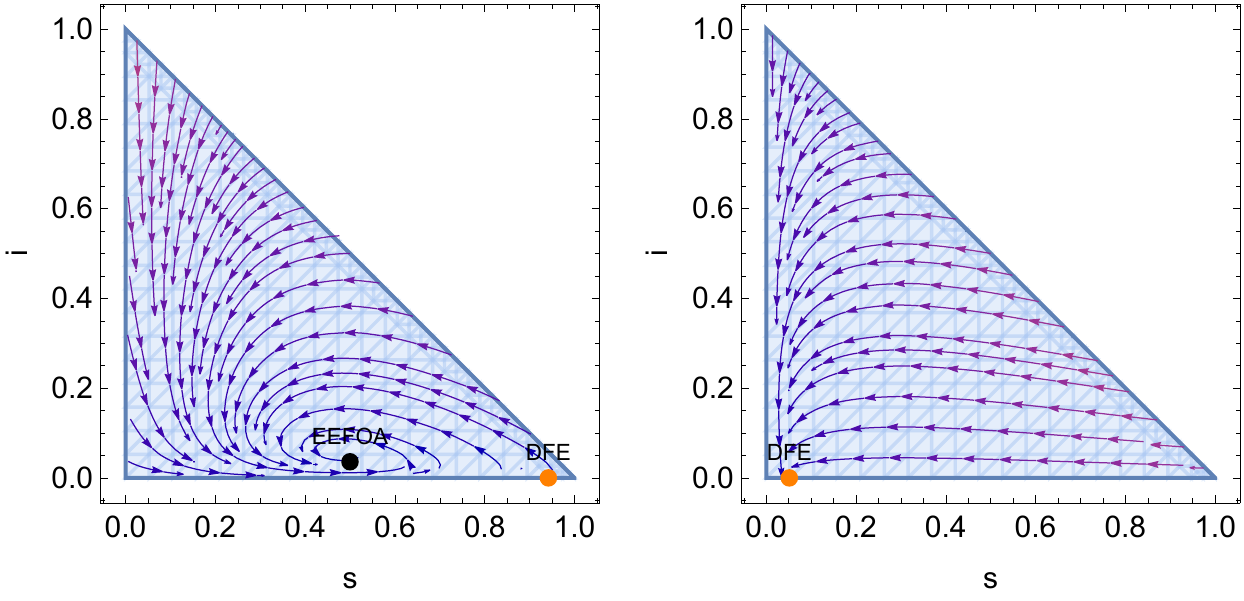}{Stream plots of $(s,i)$ for the FA model  when $\g_s \in \left\{1/100,  3\right\}$ is smaller and bigger, \resp, than the critical vaccination $\g_s^*=0.239087$ defined in \eqr{crv}.}{1}%StreamPlots.nb

\section{Appendix A: Auxiliary lemmas \la{s:lem}}
\beL  If only strict inequalities between $ \Lgn , \b_r$ and  $\b$ are allowed, then the following ten cases
$$ \bc \Lgn < \b_r< \b\\
\L+\g+\nu_i< \b< \b_r
\\\b_r < \L+\g+\nu_i < \b, \g_s <\g_s^*\\
 \b <  \L+\g+\nu_i < \b_r, \g_s \ge \g_s^*  \\ \b_r < \Lgn<\b, \nu_i<\b\\
 \b<\nu_i < \Lgn< \b_r\\
 \nu_i< \b { < \Lgn<\b_r}, \g_s<\g_s^*,\D<0\\
 \nu_i< \b { < \Lgn<\b_r}, \g_s<\g_s^*,\D\ge 0\\
 \beta <  \beta _r<  \Lgn\\\beta_r <  \beta < \Lgn\ec,$$
form a disjoint decomposition of the parameter space. \eeL
\Prf We note first that in all cases where equality is allowed, the equality case may be arbitrarily assigned to any of the two cases it separates. It is enough    therefore to consider only strict inequalities  in  this Lemma.

  We want to show that the union of the ten cases equals the union of the six cases representing the possible orders of $\b,\b_r, \Lgn$, which are
     $$\bc \beta >\beta _r>\gamma +\nu _i+\Lambda &1\\
     \beta _r>\beta >\gamma +\nu _i+\Lambda &2\\
     \beta >\gamma +\nu _i+\Lambda >\beta _r&3\\
     \beta _r>\gamma +\nu _i+\Lambda >\beta &4\\\gamma +\nu _i+\Lambda >\beta _r>\beta &5\\\gamma +\nu _i+\Lambda >\beta >\beta _r&6\ec.$$

     Now the cases $1,2,$ with $\Lgn <\min[\b,\b_r]$, and the cases $9,10,$ with $\Lgn >\min[\b,\b_e]$ appear only once in the ten cases of theorem \ref{cases}, as case 1.(a), 1(b), and 2(d)(i-ii).

     {Next, \wmc  the union of the cases 3 and 5 (1.(c) and 2.(a) in the Theorem) form together the permutation 3. This requires
      checking that the other two of the four formal cases taking into account the possible orders between $\b, \nu_i$ and $\g_s,\g_s^*$ are void;  the tedious verification is included in the Mathematica file available on
      our website}.

     To conclude, it remains to check that  the cases 4,6,7,8 (i.e. 1.(d),  2.(b), 2.(c)(i) and 2.(c)(ii) in the Theorem)  form together a partition of the permutation 4. {Note first that cases 7 and 8 may be combined in $\nu_i< \b  < \Lgn<\b_r, \g_s<\g_s^*$.} Next, we show in the Mathematica file that the case 4 $ \b <  \L+\g+\nu_i < \b_r, \g_s \ge \g_s^*$ is incompatible with $\b < \nu_i$, and so we can modify the case 4 to $ \nu_i< \b <  \L+\g+\nu_i < \b_r, \g_s \ge \g_s^*$. So, 4, 6 and 7-8 become
     $$\bc \nu_i < \b <  \L+\g+\nu_i < \b_r, \g_s \ge \g_s^*  \\
 \b<\nu_i < \Lgn< \b_r\\
 \nu_i< \b { < \Lgn<\b_r}, \g_s<\g_s^*,\ec$$
 whose union is clearly the permutation 4.\QEDB

 \beL \la{l:alg}  A) A necessary and sufficient condition for having precisely one endemic point with $s \in (0,1)$ is $C(A+B+C)<0$, where $A,B,C$ are defined in \eqr{ABCi}.

B) Necessary and sufficient conditions for having precisely two endemic points with $s \in (0,1)$ are $\D > 0 $ and
$$ -2<  \fr BA <  0 < \fr CA,    \fr BA >-1 -\fr CA.  %\Eq 0 > \fr BA >-1 -\min[1,\fr CA]\ec.
$$
\eeL

\Prf The conditions for having two roots bigger than 0 are $ \fr BA < 0 < \fr CA$, and  the conditions for having two roots smaller than 1 are obtained by applying these, after  substituting $y=1-x$, yielding  the result. \QEDB

 Since expressing these simple conditions in terms of the parameters of the model turned out quite difficult, we did not finalize  this approach.

%\iffalse
\section{Apendix B: A short review of deterministic epidemic models\la{s:pr}}
\ssec{What is a deterministic epidemic model?}

To put in perspective our work, we would like to start by a  definition of deterministic epidemic  models, lifted from \cite{KamSal}.

\beD
A deterministic epidemic  model is a  dynamical system with two types of variables $\vec x(t):=(\vi(t),\vec z(t))\in {\mathbb{R}}_{+}^{N}$, where
\BEN \im
$\vi(t)$ model the number (or  density) in
 different compartments of  infected individuals (i.e. latent,
infectious, hospitalized, etc) which should ideally disappear eventually if the epidemic ever ends;

\im $\vec z(t)$ model  numbers (or  densities) in
  compartments of individuals
who are not infected (i.e. susceptibles, immunes,
recovered individuals, etc).
\EEN

 The system must admit  an  equilibrium called \DFE\ (DFE), and hence a ``quasi-triangular" linearization the form
\beq \la{dyn} \vi'(t) &&=\vi(t) A_{i,i}(\vec x(t)), \vec x(t)\in \mD \subset{\mathbb{R}}_{+}^{N}\\
\vec z'(t)&&=\vi(t) A_{z,i}(\vec x(t)) + (\vec z(t)-\vec z_{dfe})  A_{z,z}(\vec x(t)),\no\eeq
where $\mD$  is some forward-invariant subset,
where ``quasi-triangular" refers to the fact that the functions $A_{i,i}, A_{z,i},A_{z,z}$ depend on
all the variables $\vec x(t)$, and  where $N$ is the dimension of $\vec x(t)$.
\QEDB
\eeD

As shown in  \cite{KamSal}, any epidemic model admitting an equilibrium
point  $(\vz, \vec z_{dfe})$ admits the representation \eqr{dyn}, under
suitable smoothness assumptions.
In what follows, we will call the point $x_{dfe}=(\vz, \vec z_{dfe})$ a
\DFE\  (DFE).

\beR
Note that the  essential feature of \eqr{dyn} is the ``factorization of the disease equations". %which facilitates computing the \dfe\  and the endemic equilibrium points.
\QEDB
\eeR

\subsection{The \brn\ and the \ngm\ method}

One of the central objectives in mathematical epidemiology
is to study the stability of DFE, i.e. the conditions which make possible eradicating the sickness. For simple models, this amounts, independently of the initial state,  to verifying that a famous threshold parameter called \brn\ $\m0$ is less than $1$ (and  the fundamental  problem of interest is,  when  $\m0>1$, to offer control strategies
which force the epidemic to reach the DFE).

 {The \brn\ or ``net reproduction rate" $\m0$  is a pillar concept in demography, branching processes and mathematical epidemiology -- see the introduction of the book \cite{bacaer2021mathematiques}.}
 \BEN \im The notation was  first introduced by the father
of mathematical demography Lotka \cite{lotka1939analyse,dietz1993estimation}. In epidemiology, the basic reproduction number models
 the expected number of secondary cases which one infected case would produce in a  homogeneous,
completely susceptible stochastic population, in the next generation.  As well known in the simplest setup of branching  process, having this  parameter smaller than $1$ makes extinction sure.  The relation to epidemiology is   that an epidemics is well approximated by a branching process at its inception, a fact which goes back to  Bartlett and Kendall.%\cite{kendall2020deterministic}.
\im With more infectious classes, one deals at inception with  multi-class branching processes, and stability holds when the \PF\ eigenvalue of the   ``{\bf \ngm\ }" (NGM)  of means  is smaller than $1$.

  \im   For  deterministic epidemic models, it seems at first that the \brn\ $\m0$ is lost, since the  generations disappear in this setup -- but see  \cite[Ch. 3]{bacaer2021mathematiques}, who recalls a method to introduce generations which goes back to Lotka, and which is reminiscent of the iterative Lotka-Volterra approach of solving integro-differential. At the end of the tunnel, a unified method for defining $\m0$ emerged only much later, via
      the ``next generation matrix"  approach \cite{diekmann1990definition,Van,Van08,Diek,perasso2018introduction}.
The final result is that local stability of the \DFE\ holds iff the spectral radius of a certain matrix called ``\ngm",  which depends only on a set  of  ``infectious  compartments" $\vi$ (which we aim to   reduce to $0$), is less than one.
This  approach works provided that certain  assumptions listed below hold\fn[4]{And so $\m0$ is undefined when these assumptions are not \satd.}.
\EEN

      \BEN \im The foremost assumption is that the disease-free  equilibrium $(\vec 0, z_{dfe})$ is
{\em unique and locally asymptotically stable within the disease-free space} $\vi=0$, meaning that all solutions of   $$\vec z'(t)=(\vec z(t)-\vec z_{dfe})  A_{z,z}(\vz,\vec z(t)), \qu \vec z(0)=\vec z_0$$
 must approach the point $z_{dfe}$  when $t\to \I$).

\im  Other conditions are related  to an ``admissible splitting" as a difference of two  parts $\mF,\mV$, called \resp\  ``new infections", and  ``transitions"
\beD A splitting
$$\vi'(t)=\mF(\vi(t),\vec z(t))-\mV(\vi(t),\vec z(t))$$
will be called admissible if
$\mF,\mV$  \saty\ the  following  conditions  \cite{Van08,Shuai}:
\BEN \im
\be{cond} \bc \mathcal{F}(\vec 0,\vec z(t))=  \mathcal{V}(\vec 0,\vec z(t))= 0\\
\mathcal{F}(\vi(t),\vec z(t))\geq 0, \for (\vi(t),\vec z(t)) \\
\mathcal{V}_j(\vi(t),\vec z(t))\le 0, \text{ when } \vi_j=0,\\
\sum_{j=1}^n \mathcal{V}_j(\vi(t),\vec z(t))\geq 0, \for (\vi(t),\vec z(t)).
\ec,\ee
where the subscript $j$ refers the $j$'th component.
\EEN
\eeD

\beR The splitting of the infectious equations  has its origins in epidemiology. Mathematically, it is related to the "splitting of Metzler matrices"-- see \fe\ \cite{fall2007epidemiological}.
Note however  that the splitting conditions may  be satisfied for several or for no subset of \com s
(see \fe\ the SEIT model, discussed in \cite{Van08}, \cite[Ch 5]{Mart}). Unfortunately, for deterministic epidemic models,  there is no   clear-cut definition of $\m0$ \cite{roberts2007pluses,li2011failure,Thieme}.\fn[3]{A possible explanation is that several stochastic epidemiological  models may correspond in the limit
to the same deterministic  model.}\eeR

\im We turn now to the last conditions, which concern the linearization of the infectious equations around the DFE. Putting
 $L =A_{i,i}(\vz, \vec z_{dfe})$, and letting $f$ denote the perturbation from the linearization, \wmw:
\beq \la{dynl} \vi'(t) &&=\vi(t) L-f(\vi(t),\vec z(t))=\vi(t)(F-V)-
f(\vi(t),\vec z(t)),\\&&  F:=\left[\frac{\partial \mathcal{F}}{\partial \i}\right]_{x_{dfe}}
 V=\left[\frac{\partial \mathcal{V}}{\partial \i}\right]_{x_{dfe}}, L=F-V. \no \eeq

The ``transmission and transition" linearization matrices  $F,V$
must \saty\ that $F\ge 0$ componentwise and   {$V$ is a non-singular M-matrix}, which ensures that $V^{-1}\ge 0$.\fn[6]{The  assumption (B) implies that   $L=F-V$ is a ``stability (non-singular) M-matrix", which is necessary for the non-negativity and boundedness of the solutions \cite[Thm. 1-3]{de2019some}.}

 \EEN

Under the  assumptions above, the \ngm\ method concludes that $\m0:=\l_{PF}(FV^{-1})$ is a threshhold parameter in the following sense  \cite[Thm 1]{Van08}:
 \BEN \im
\BEN \im When $\m0<1$,
 {the DFE is locally asymptotically stable},   while
 \im when $\m0>1$, the DFE is unstable.
  \EEN
  We will call this the ``weak $\m0$ alternative".

  \im Also, global asymptotic stability of the DFE holds  when $\m0 \le 1$, provided the ``perturbation from linearity"
  $f=i(F-V) - \mF +\mV$ is non-negative \cite[Thm 2]{Van08}.
 This result has been called the ``strong $\m0$ alternative" in \cite{guo2006global,Shuai}.
\fn[5]{Note that the \cite{guo2006global} strong $\m0$ alternative was only established for a general n-stage-progression, which is a particular case of the model we study below, in which  A is an ``Erlang" upper diagonal matrix. It is natural to expect that the result continues to hold for other  non-singular Metzler matrices}.
\EEN
\iffalse
   Note also that  for the ``bilinear incidence" case with infection rate $\b S I$,  when $\m0>1$, a unique endemic point is often the global attractor, a phenomena which does not happen already in the first example \cite{Der} studied for the next class of models.

  \fi

\iffalse
\fi

\section*{Acknowledgments}
We thank N. Bacaer for providing the references \cite{bacaer2021mathematiques,lotka1939analyse}, and we thank the referees for their thorough reviews and suggestions.
%
%\reftitle{References}
\bibliographystyle{amsalpha}

\bibliography{Pare38}

\end{document}